\documentclass[11pt]{article}
\usepackage{amsmath,amssymb}
\usepackage{graphicx}

\renewcommand{\baselinestretch}{1.5}

\makeatletter
\@addtoreset{equation}{section}
\makeatother

\newcommand{\R}{{\mathbb{R}}}
\newcommand{\Z}{{\mathbb{Z}}}

\DeclareMathOperator{\Tr}{Tr}
\DeclareMathOperator{\tr}{tr}
\newcommand{\bhline}{\noalign{\hrule height 1pt}}

\newcounter{dummy}
\newcommand{\twoB}{\hbox{\Roman{dummy}B} }
\newcommand{\omc}{\omega} 
\newcommand{\om}{\theta}  

\begin{document}

\begin{flushright}
 \sf
 hep-th/9903233\\
 UTMS 99-16\\
 OCHA-PP135
\end{flushright}

\vfill

\begin{center}
 {\huge Comments on Large $N$ Matrix Model } 
 
 \vspace*{2.0cm}
 
 {\large 
 Hiroshige Kajiura$^{a\dagger}$, ~
 Akishi Kato$^{a\star}$, ~ 
 Sachiko Ogushi$^{b\ast}$ }
 
 \vspace*{1.0cm}
 
 ${}^{a)}\; ${\large \it 
 Graduate School of Mathematical Sciences, University of Tokyo \\
 Komaba 3-8-1, Meguro-ku, Tokyo 153-8914, Japan}\\
 ${}^{b)}\; ${\large \it Department of Physics,
 Faculty of Science, Ochanomizu University \\
 Otsuka 2-1-1, Bunkyo-ku, Tokyo 112-8610, Japan }\\
 
 \thispagestyle{empty}
 \vspace*{1.0cm}
 
 {\bf Abstract}\\
 
 
 \vspace*{0.3cm}
 
 \begin{minipage}[t]{.88\textwidth}
  The large $N$ Matrix model is studied with attention to the quantum
  fluctuations around a given diagonal background. Feynman rules are
  explicitly derived and their relation to those in usual Yang-Mills
  theory is discussed. Background D-instanton configuration is naturally
  identified as a discretization of momentum space of a corresponding
  QFT.  The structure of large $N$ divergence is also studied on the
  analogy of UV divergences in QFT.
  
  
  
 \end{minipage}
 
\end{center}

\vfill

\begin{flushleft}
 \renewcommand{\baselinestretch}{1.0}
 \rule{100mm}{.2pt}\\
 \begin{small}
  \sf
  $\dagger$ kuzzy@ms.u-tokyo.ac.jp\\
  $\star$ akishi@ms.u-tokyo.ac.jp\\
  $\ast$ g9740505@edu.cc.ocha.ac.jp\\
 \end{small}
\end{flushleft}
\newpage


\setcounter{page}{1}

 \section{Introduction}
\label{sec:Introduction}

Matrix model provides a new paradigm for thinking about fundamental
theories of physics. It originates in the observation \cite{Witten} that
the massless modes propagating along the world volume of $N$ coincident
D-branes are those of the supersymmetric Yang-Mills theory, obtained by
the dimensional reductions of the $D=10$ $N=1$ theory down to $p+1$
spacetime dimensions.

According to so-called Matrix conjecture \cite{BFSS}, $0+1$-dimensional
reduction can be regarded as the discrete light cone quantization of
M-theory in which the spacetime is compactified on an almost light like
circle. This proposes a concrete, a nonperturbative definition of
quantum gravity, and quite remarkably, the conjecture has found quite
nontrivial support \cite{Becker,DKPS,BBPT,OY}.

Meanwhile, type \twoB matrix model proposed by \cite{IKKT} plays
somewhat complementary role.  Its action is $0+0$-dimensional reduction
of large $N$ super Yang-Mills theory in ten dimensions. The authors of
\cite{AIKKT} proposed a very interesting program to study dynamical
formation of space-time using the type \twoB matrix model.

Despite much remarkable success of Matrix model approach, the question
``Why and how such a simple model could describe our real world?'' is
still elusive. The main difficulty consists in the absence of built-in
rules concerning ``How to take large $N$ limit.''  For example, in the
case of Matrix model approach to 2d gravity \cite{OldMatrix}, there is a
critical point $g_{c}$, and continuum limit is possible keeping certain
relation between $N$ and $g-g_{c}$ (double scaling limit).  For type
\twoB matrix model, however, the coupling constant $g$ can be absorbed
into the rescaling of the fields (at least classically) and there is no
nontrivial fixed point.

The result of matrix integration is just a number as it stands.  To
extract physical intuition, we need to separate field variables into two
types: the classical background and the quantum fluctuation.

In the spirit of Born-Oppenheimer approximation, the effective dynamics
of the slow variables (classical background) are of primary concern
which is obtained only after fast variables (quantum fluctuations) are
integrated out.  This is the approach taken by many works.

In Matrix models, however, somewhat different approach might be of
considerable interest. Recall that in the usual analysis of quantum
field theory, gravitational effects are almost always ignored, although
gravitons are massless and never decouple.  Gravitational degree of
freedom are not integrated over, but regarded as fixed, classical
background. This treatment is justified simply because the dimensionful
coupling is so small in the energy scale accessible by the current
technology.  Similarly for the observers living on the branes, natural
time scale is set by that of quantum fluctuations rather than the
dynamical time scale of the background = spacetime. Put it differently,
``motion of the background is too slow to be treated quantum
mechanically.''

In this paper, we will study the quantum dynamics of the Matrix model
from the latter point of view, hoping our work provide some insight
about how to take large $N$ limit. The paper is organized as follows.

In section two, starting from $0+0$-dimensional matrix action, we derive
fatgraph Feynman rules for the quantum fluctuations treating general
multi D-instanton configuration as a fixed background. The usage of the
Feynman rules is shown with an example. Although we will work in
D-instanton backgrounds of \twoB matrix model, we expect our analysis
shed some light on general D-$p$ branes in other Matrix theories as
well, since type \twoB matrix model compactified on $S^{1}$ is
equivalent to the $1+0$-dimensional Matrix model \cite{CDS}.

The matrix Feynman rules are very close to those in the usual
$d$-dimensional SYM.  In section three, we will study a special
backgrounds where D-instantons are concentrated along $d$-dimensional
sheet in the original $D$-dimensional spacetime. We will see finite $N$
theories can be thought of as UV regulated versions of flat space
Yang-Mills theory in which removing the cutoff is equivalent to letting
$N$ go to infinity.  The crucial observation of this paper is that from
Yang-Mills perturbation point of view, going to Matrix model can be
thought of as a discretization of a momentum space rather than a
coordinate space.  This is shown explicitly by comparing Feynman
rules. This is yet another manifestation of spacetime uncertainty
\cite{Yoneya} or UV/IR correspondence \cite{SW}.

For the $d$-dimensional quantum field theory embedded in the Matrix
model, the only source of divergence is the large $N$ limit.  In section
four, we will study the structure of large $N$ divergences in Matrix
theory and relate it to the renormalizability of QFT in the usual sense
of the term.  We hope this line of argument give us a hint to deduce
realistic physics from the Matrix models.  Finally, we conclude with a
discussion of our results and some implications.

 \section{Matrix perturbation theory around D-instanton background}
\label{sec:Perturbation}

In this section, we elaborate the perturbation theory of the Matrix
model around D-instanton background. Explicit forms of Feynman rules are
derived and boson self energy diagrams are computed at one loop as an
example.

  \subsection{The type \twoB matrix model and its gauge fixing}

Our starting point is the Euclidean type \twoB matrix model, whose
action is given by
\begin{equation}
 S = -\frac{1}{g^2}\Tr
  \left(
   \frac{1}{4}[X_{\mu},X_{\nu}]^{2} 
   + \frac{1}{2}\bar{\psi }\Gamma ^{\mu}
   [X_{\mu},\psi] \right) \label{saisyo}
\end{equation}
where $X_{\mu}$ and $\psi$ are $D$-dimensional vector and Majorana-Weyl
spinor respectively, taking values in $N \times N$ hermitian
matrices.\footnote{We choose $D=10$ type \twoB matrix model just for
definiteness.  We could start from any model reduced from
$D$-dimensional SYM.}  Throughout this paper, ``Tr'' denotes the trace
taken over $N \times N$ matrix indices.

The action enjoys the following symmetries
\begin{itemize}
 \itemsep=-2.5ex
 \item {rotation invariance}
      \begin{equation*}
       \delta X^{\mu} = \omega ^{\mu\nu} X^{\nu}, \qquad
	\delta \psi = \frac{i}{2}\omega^{\mu\nu}\Gamma ^{\mu\nu}\psi,
	\qquad (\omega ^{\mu\nu}=-\omega^{\nu\mu})
      \end{equation*}
 \item {translation invariance}
      \begin{equation}
       \delta X^{\mu}= c^{\mu}
	\label{translation}
      \end{equation}
 \item {${\cal N}=2$ SUSY}
      \begin{equation*}
       \delta^{(1)}\psi = \frac{i}{2}
	[X_{\mu},X_{\nu}]\Gamma^{\mu\nu}\epsilon _1,\qquad
	\delta^{(1)} X_{\mu} = i\bar{\epsilon _1}\Gamma_{\mu}\psi ,
	\qquad \delta^{(2)}\psi = \epsilon _2 ,\qquad
	\delta^{(2)} X_{\mu} = 0.
      \end{equation*} 
 \item {$U(N)$ Gauge invariance}
      \begin{equation}
       X^{\mu} \mapsto U^{-1} X^{\mu} U, \qquad
        \psi \mapsto U^{-1} \psi^{\mu} U, \qquad
        (U \in U(N))
        \label{GaugeAction}
      \end{equation}
 \item {scaling property :} 
      \begin{equation}
       X^{\mu} \rightarrow \lambda X^{\mu}, \qquad
        g \rightarrow \lambda^2 g
        \qquad (\lambda \in \R_{>0})
        \label{dimension}
      \end{equation}
\end{itemize}

As stated in Introduction, we decompose $X^{\mu}$ as a sum of classical
background part $\bar{X}^{\mu}$ and quantum fluctuation part
$\tilde{X}^{\mu}$. $\bar{X}^{\mu}$ will be treated as fixed, classical
number and we will be interested in the quantum field theory in this
background.  ($\psi$ is assumed to have no classical vacuum expectation
value.)

The background $\bar{X}^{\mu}$ must be a solution to the equation of
motion, $[X^{\mu},[X^{\nu},X^{\mu}]]=0$.  We will consider the cases
where all the $\bar{X}^{\mu}$'s are simultaneously diagonalizable by the
gauge action (\ref{GaugeAction}):
\begin{equation}
 X^{\mu} = \bar{X}^{\mu}+\tilde{X}^{\mu},
  \qquad
  \bar{X}^{\mu} 
  \equiv \left( 
          \begin{array}{llll} x_{1}^{\mu} & & & \\ & x_{2}^{\mu} & & \\
	   & & \ddots & 
          \\ & & & x_N^{\mu} \end{array} \right) . 
   \label{waketa} 
\end{equation}
The combination of $D$ eigenvalues $x_{i} \equiv (x_{i}^{1},\cdots
x_{i}^{D}) \in \R ^{D} $ is interpreted as the location of the $i$-th
D-instanton. For a generic background where all D-instantons are
separated from each other, all the symmetries listed above are
explicitly broken.  In particular, $U(N)$ gauge symmetry is broken down
to $U(1)^{N}$ and half of the ${\cal N}=2$ SUSY (generated by
$\delta^{(1)}$) survives indicating the BPS nature of the background
(\ref{waketa}).

Let us make a brief comment on the charges of the fields.  All the
quantum fields $\tilde{X}_{ij}^{\mu},\psi _{ij},c_{ij}$, $b_{ij}$ have a
charge
\begin{equation}
 \renewcommand{\arraystretch}{.4}
  \arraycolsep=2pt
  \begin{array}{ccccccccccccc}
   &  &       &  &i      &  &       &  &j      &  &       &  &  \\
   &  &       &  &\smile &  &       &  &\smile &  &       &  &  \\
 ( &0 &\cdots &0 &1      &0 &\cdots &0 &-1     &0 &\cdots &0 &) \\
  \end{array}
  \label{chargevector}
\end{equation}
with respect to the unbroken $U(1)^{N}$ gauge symmetry. These
fluctuations correspond to the open string stretching between
D-instantons $i$ and $j$.  In particular, diagonal components
$\tilde{X}_{ii}^{\mu},\psi_{ii},c_{ii}$ and $b_{ii}$ are neutral.  In
fact, as we will soon see, their kinetic terms vanishe indicating they
should be treated as collective coordinates rather than quantum
variables, and thus need a separate treatment. Since these diagonal
components could be absorbed into the shift of the background D-instanton
configuration, incorporating these fluctuations would inevitably lead to
the integral over the collective coordinates, which is beyond the scope
of this paper.

We will study the quantum theory of fluctuations as parameterized by the
classical background i.e. D-instanton positions $\{x_{i}\}_{i=1}^{N}$.
Plugging (\ref{waketa}) into the action (\ref{saisyo}) and using the
relation $[\bar{X}^{\mu},\tilde{X}^{\nu}]_{ik}=(x_{i}^{\mu}-x_{k}^{\mu})
\tilde{X}_{ik}^{\nu}$, we have
\begin{eqnarray}
 S &=& \frac{1}{2g^2}\sum_{i,j}\{(x_{ij})^2
  \tilde{X}_{ij}^{\nu}\tilde{X}_{ji}^{\nu}\} 
  + \frac{1}{2g^2}\sum_{i,j}\{ x_{ij}^{\mu}x_{ji}^{\nu}
  \tilde{X}_{ij}^{\mu}\tilde{X}_{ji}^{\nu}\} \nonumber \\
 & & +\frac{1}{g^2}\sum_{i,j,k}\{(x_{ik}+x_{jk})^{\mu}
  \tilde{X}_{ij}^{\mu}\tilde{X}_{jk}^{\nu}\tilde{X}_{ki}^{\nu} \} 
  + \frac{1}{2g^2}\sum_{i,j,k,l}\{\tilde{X}_{ij}^{\mu}\tilde{X}_{jk}^{\mu}
  \tilde{X}_{kl}^{\nu}\tilde{X}_{li}^{\nu} 
  - \tilde{X}_{ij}^{\mu}\tilde{X}_{jk}^{\nu}
  \tilde{X}_{kl}^{\mu}\tilde{X}_{li}^{\nu}\}
  \nonumber \\
 & & -\frac{1}{2g^{2}}\sum_{i,j}\{ \bar{\psi}_{ji}
  \Gamma^{\mu}x_{ij}^{\mu}\psi_{ij}\}
  +\frac{1}{2g^{2}}\sum_{i,j,k}\{\bar{\psi}_{ij}
  \Gamma^{\mu}\psi_{jk}\tilde{X}_{ki}^{\mu}
  -\bar{\psi}_{ij}\Gamma^{\mu}\tilde{X}_{jk}^{\mu}\psi_{ki} \}.
  \label{tenkai}
\end{eqnarray}
It should be noted that $x_{i}$'s always appear as difference
$x_{i}-x_{j}$ due to translational invariance (\ref{translation}).
Hereafter the notation $x_{ij}^{\mu}\equiv x_{i}^{\mu}-x_{j}^{\mu}$ will
be used to simplify the formulas.

To setup a perturbation theory, convenient to work with the background
 field gauge.  This is achieved by adding the gauge fixing term
\begin{equation}
 S_{g.f.} = -\frac{1}{2g^2}\Tr[\bar{X}^{\mu},\tilde{X}^{\nu}]^{2}
  =-\frac{1}{2g^2}\sum_{i,j}\{x_{ij}^{\mu}
  x_{ji}^{\nu}\tilde{X}_{ij}^{\mu}\tilde{X}_{ji}^{\nu}\}
  \label{Sgf}
\end{equation}
accompanied with the Faddeev-Popov ghost term
\begin{eqnarray}
 S_{F.P.} &=& -\frac{1}{g^2}\Tr[\bar{X}^{\mu},b]\,[X^{\mu},c] \nonumber \\
 &=& \frac{1}{g^{2}}\sum_{i,j}\{(x_{ij}^{\mu})^{2}b_{ij}c_{ji} \}
  +\frac{1}{g^{2}}\sum_{i,j,k}\{x_{ij}^{\mu}b_{ij}
  (c_{jk}\tilde{X}_{ki}^{\mu}-\tilde{X}_{jk}^{\mu}c_{ki})\}
  \label{ghost}.
\end{eqnarray}
The gauge fixing term (\ref{Sgf}) implies we have chosen a gauge such
that the fluctuation $\tilde{X}_{ij}$ is transverse to the relative
vector $x_{ij}$, i.e. $\sum_{\mu}x_{ij}^{\mu}\tilde{X}_{ij}^{\mu}=0$.

The gauge fixed total action is given by
\begin{equation}
 S_{total} =S+S_{g.f.}+S_{F.P.}
  \label{total}
\end{equation}
Note that $S_{g.f.}$ in (\ref{total}) cancels with the second term of
(\ref{tenkai}).  
The perturbation is valid when D-instanton separation $x_{ij}$ is much
larger than $g^{1/2}$.

  \subsection{Feynman rules}
\label{sec:Feynman}

Now we will derive Feynman rules from the gauge fixed action
(\ref{total}). Hereafter we will rescale the quantum fluctuations
$\tilde{X}_{ij}^{\mu} \to g \tilde{X}_{ij}^{\mu}$, etc, in order that
the coupling $g$ is removed from the propagators and moved to the
interaction vertices.

\begin{figure}[tbp]
 \centering
 \includegraphics*[width=.5\textwidth,keepaspectratio=true]{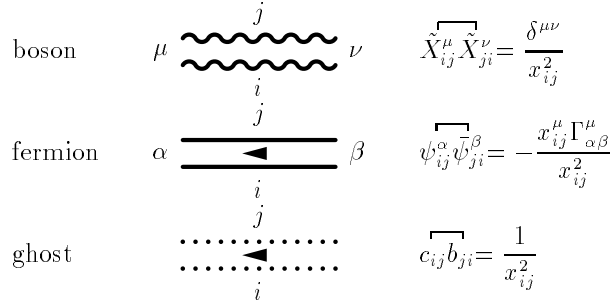}
 \caption{Propagators}\label{fig:propagator}
\end{figure}

\begin{figure}[tbhp]
 \centering
 \includegraphics*[width=.7\textwidth,keepaspectratio=true]{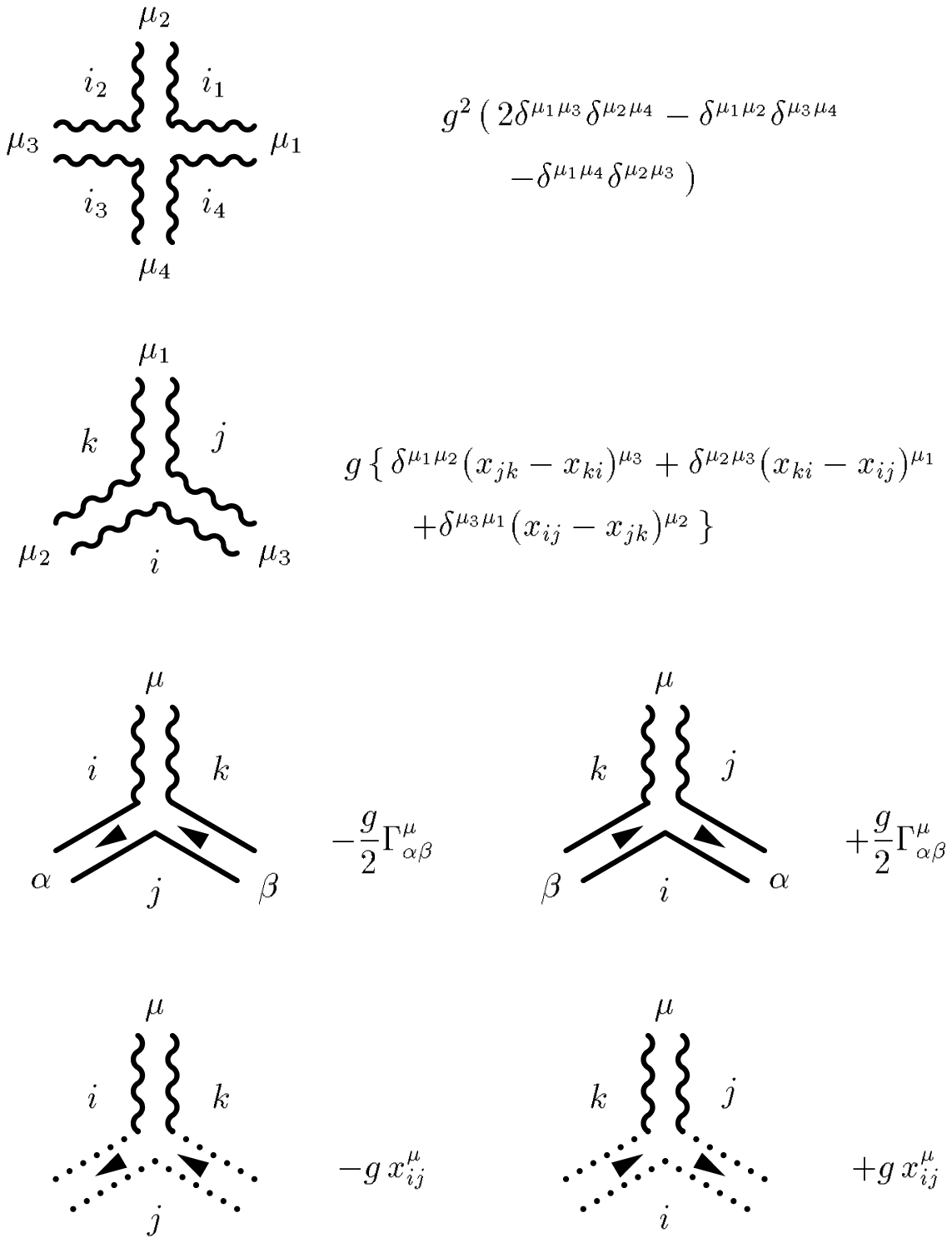}
 \caption{Vertices}\label{fig:vertex}
\end{figure}

The perturbative structure of large $N$ gauge theories are naturally
described in terms of double line representation \cite{tHooft} of
Feynman diagrams, so called fatgraphs.  One considers, as in Figs.
\ref{fig:propagator} and \ref{fig:vertex}, a graph with the lines
thickened slightly into bands which meet smoothly at the vertices so as
to form an oriented Riemann surface with boundary.  In this
representation, the fields in the adjoint representation of $U(N)$ are
denoted by double lines.

Each edge carries a label $i,j,\ldots \in \{1,\ldots,N\}$ corresponding
to the basis of fundamental representation of $U(N)$ (or its conjugate
depending on its orientation).  Here it is nothing but the label of a
D-instanton.  From the quadratic part of the total action (\ref{total}),
we can easily read off the fatgraph propagators as depicted in
Fig.~\ref{fig:propagator}. Note that the denominator is the squared
distance between the two D-instantons connected by the fields.

As for vertices, there is a crucial difference between ordinary
(particle theory) vertex factor and fatgraph counterpart. In the former,
interaction vertex of order $k$ is invariant under all possible $k!$
permutations of lines, while in the latter, it is invariant only under
$k$ cyclic permutations.  Thus, for example, two Yukawa coupling
diagrams in Fig.~\ref{fig:vertex} should be distinguished from each
other.

In deriving the vertex factors, terms must be organized so that the
index contraction should have manifest cyclic invariance. For instance,
in order to deduce three-point and four-point vertex for $\tilde{X}$'s,
we must rewrite corresponding terms in (\ref{tenkai}) as follows
\begin{eqnarray}
 \lefteqn{
  \frac{1}{g^2}\sum_{i,j,k}\{(x_{ik}+x_{jk})^{\mu}
  \tilde{X}_{ij}^{\mu}
  \tilde{X}_{jk}^{\nu}\tilde{X}_{ki}^{\nu} \} }
  \nonumber \\  
 & = &   
  -\frac{1}{3 g^2}\sum_{i,j,k}
  \sum_{\mu_{1},\mu_{2},\mu_{3}}
  \tilde{X}_{jk}^{\mu_{1}}\tilde{X}_{ki}^{\mu_{2}}
  \tilde{X}_{ij}^{\mu_{3}}
  \nonumber  \\
 & & 
  \qquad \times 
  \left\{ 
   \delta^{\mu_{1}\mu_{2}}(x_{jk}-x_{ki})^{\mu_{3}}
   +\delta^{\mu_{2}\mu_{3}}(x_{ki}-x_{ij})^{\mu_{1}}
   +\delta^{\mu_{3}\mu_{1}}(x_{ij}-x_{jk})^{\mu_{2}}\right\} 
    \nonumber 
\end{eqnarray}
\begin{eqnarray}
 \nonumber \\
 \lefteqn{
  \frac{1}{2g^2}\sum_{i,j,k,\ell }
  \{\tilde{X}_{ij}^{\mu}\tilde{X}_{jk}^{\mu}
  \tilde{X}_{k\ell }^{\nu}\tilde{X}_{\ell i}^{\nu}- 
  \tilde{X}_{ij}^{\mu}\tilde{X}_{jk}^{\nu}
  \tilde{X}_{k\ell }^{\mu}\tilde{X}_{\ell i}^{\nu}\} 
  }
  \nonumber \\
 &=& - \frac{1}{4g^{2}}\sum_{i,j,k,\ell}
  \sum_{\mu_{1},\cdots ,\mu_{4}}
  \tilde{X}^{\mu_{1}}_{\ell i}\tilde{X}^{\mu_{2}}_{ij}
  \tilde{X}^{\mu_{3}}_{jk}\tilde{X}^{\mu_{4}}_{k\ell} 
  \nonumber \\
 & & \qquad \times 
  \left\{ 2\delta^{\mu_{1}\mu_{3} }\delta^{\mu_{2}\mu_{4}}
   -\delta^{\mu_{1}\mu_{2}}\delta^{\mu_{3}\mu_{4}}
   -\delta^{\mu_{1}\mu_{4} }\delta^{\mu_{2}\mu_{3}} \right\}. \nonumber
\end{eqnarray}
Overall factors $1/3$ and $1/4$ are cancelled by the cyclic symmetry of
the vertices.

Similar computation leads to the vertex factors listed in
Fig.~\ref{fig:vertex}.

  \subsection{Example: one loop boson self energy}
\label{sec:oneloop}

\begin{figure}\centering
  \includegraphics*[width=.85\textwidth,keepaspectratio=true]{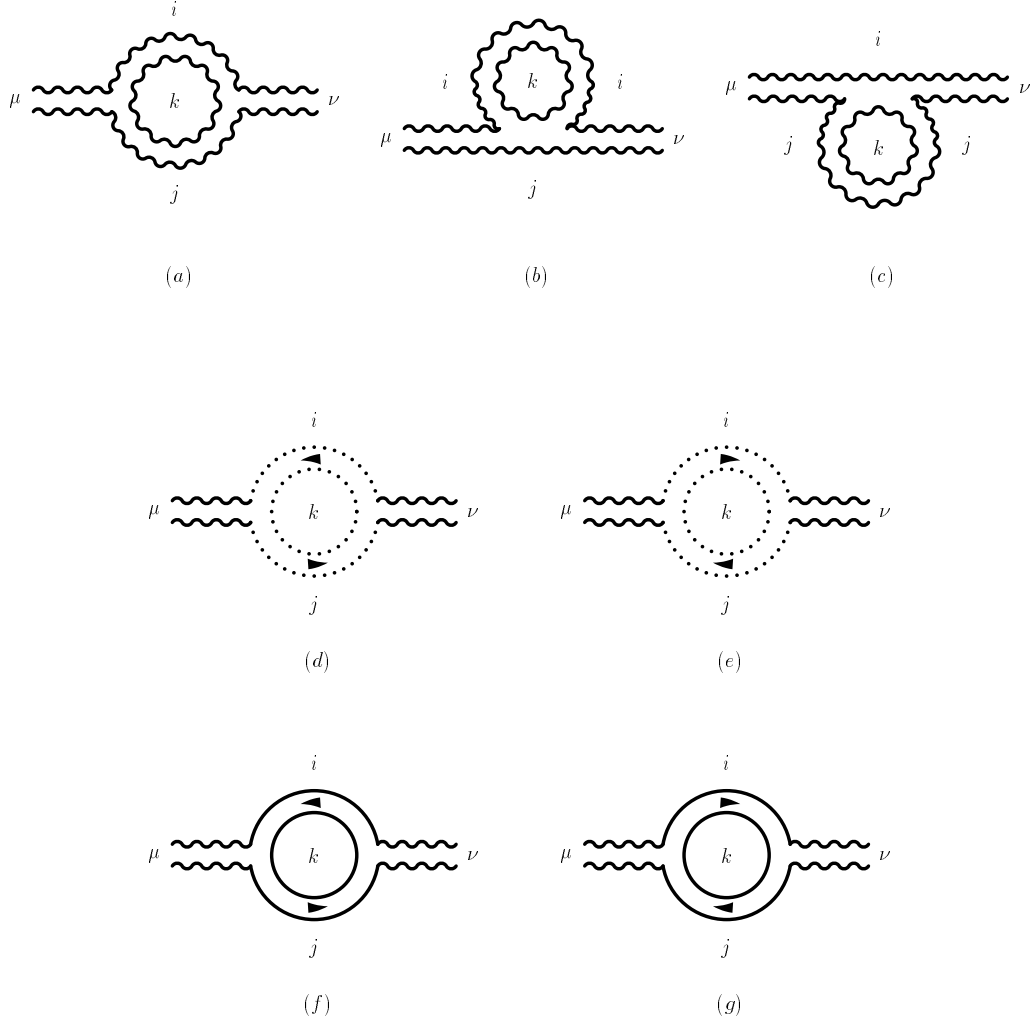}
  \caption{One loop boson self energy graphs}\label{fig:oneloop}
\end{figure}

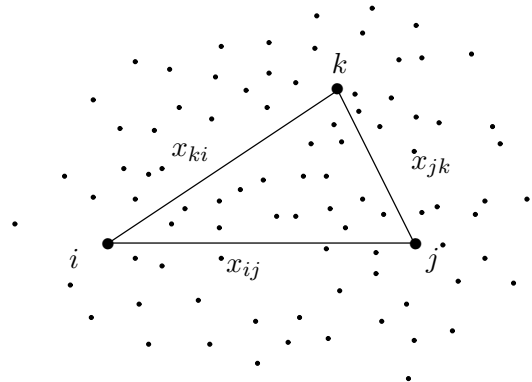
\begin{figure}[htbp]
 \begin{center}
  \unitlength 0.1in
  \begin{picture}(26.80,19.40)(1.10,-27.10)
   %
   \special{pn 8}%
   \special{pa 600 2000}%
   \special{pa 1800 1200}%
   \special{fp}%
   \special{pa 600 2000}%
   \special{pa 2200 2000}%
   \special{fp}%
   \special{pa 1800 1200}%
   \special{pa 2200 2000}%
   \special{fp}%
   \put(9.3000,-15.7000){\makebox(0,0)[lb]{$x_{ki}$ }}%
   \put(22.9000,-16.0000){\makebox(0,0){$x_{jk}$}}%
   \put(13.1000,-21.4000){\makebox(0,0){$x_{ij}$}}%
   \put(18.1000,-10.6000){\makebox(0,0){$k$}}%
   \put(23.0000,-20.9000){\makebox(0,0){$j$}}%
   \put(4.2000,-20.8000){\makebox(0,0){$i$}}%
   \put(5.6000,-20.4000){\makebox(0,0)[lb]{$\bullet$}}%
   \put(17.6000,-12.3000){\makebox(0,0)[lb]{$\bullet$}}%
   \put(21.7000,-20.4000){\makebox(0,0)[lb]{$\bullet$}}%
   %
   \special{pn 8}%
   \special{sh 1}%
   \special{ar 760 2320 10 10 0  6.28318530717959E+0000}%
   \special{sh 1}%
   \special{ar 1070 2300 10 10 0  6.28318530717959E+0000}%
   \special{sh 1}%
   \special{ar 400 2220 10 10 0  6.28318530717959E+0000}%
   \special{sh 1}%
   \special{ar 110 1900 10 10 0  6.28318530717959E+0000}%
   \special{sh 1}%
   \special{ar 510 2390 10 10 0  6.28318530717959E+0000}%
   \special{sh 1}%
   \special{ar 1130 2530 10 10 0  6.28318530717959E+0000}%
   \special{sh 1}%
   \special{ar 1370 2410 10 10 0  6.28318530717959E+0000}%
   \special{sh 1}%
   \special{ar 2030 2410 10 10 0  6.28318530717959E+0000}%
   \special{sh 1}%
   \special{ar 2430 2200 10 10 0  6.28318530717959E+0000}%
   \special{sh 1}%
   \special{ar 2570 1780 10 10 0  6.28318530717959E+0000}%
   \special{sh 1}%
   \special{ar 2650 1480 10 10 0  6.28318530717959E+0000}%
   \special{sh 1}%
   \special{ar 2790 1770 10 10 0  6.28318530717959E+0000}%
   \special{sh 1}%
   \special{ar 2570 2080 10 10 0  6.28318530717959E+0000}%
   \special{sh 1}%
   \special{ar 840 1410 10 10 0  6.28318530717959E+0000}%
   \special{sh 1}%
   \special{ar 1980 770 10 10 0  6.28318530717959E+0000}%
   \special{sh 1}%
   \special{ar 1690 810 10 10 0  6.28318530717959E+0000}%
   \special{sh 1}%
   \special{ar 1320 1200 10 10 0  6.28318530717959E+0000}%
   \special{sh 1}%
   \special{ar 1410 1670 10 10 0  6.28318530717959E+0000}%
   \special{sh 1}%
   \special{ar 1580 1860 10 10 0  6.28318530717959E+0000}%
   \special{sh 1}%
   \special{ar 1540 2520 10 10 0  6.28318530717959E+0000}%
   \special{sh 1}%
   \special{ar 1380 2630 10 10 0  6.28318530717959E+0000}%
   \special{sh 1}%
   \special{ar 2170 2710 10 10 0  6.28318530717959E+0000}%
   \special{sh 1}%
   \special{ar 2400 2460 10 10 0  6.28318530717959E+0000}%
   \special{sh 1}%
   \special{ar 2200 2510 10 10 0  6.28318530717959E+0000}%
   \special{sh 1}%
   \special{ar 810 2530 10 10 0  6.28318530717959E+0000}%
   \special{sh 1}%
   \special{ar 370 1650 10 10 0  6.28318530717959E+0000}%
   \special{sh 1}%
   \special{ar 2160 2320 10 10 0  6.28318530717959E+0000}%
   %
   \special{pn 8}%
   \special{sh 1}%
   \special{ar 1430 1290 10 10 0  6.28318530717959E+0000}%
   \special{sh 1}%
   \special{ar 1580 1160 10 10 0  6.28318530717959E+0000}%
   \special{sh 1}%
   \special{ar 1190 970 10 10 0  6.28318530717959E+0000}%
   \special{sh 1}%
   \special{ar 1140 1350 10 10 0  6.28318530717959E+0000}%
   \special{sh 1}%
   \special{ar 1930 1770 10 10 0  6.28318530717959E+0000}%
   \special{sh 1}%
   \special{ar 1840 1920 10 10 0  6.28318530717959E+0000}%
   \special{sh 1}%
   \special{ar 1740 1650 10 10 0  6.28318530717959E+0000}%
   \special{sh 1}%
   \special{ar 1620 1650 10 10 0  6.28318530717959E+0000}%
   \special{sh 1}%
   \special{ar 1820 1470 10 10 0  6.28318530717959E+0000}%
   \special{sh 1}%
   \special{ar 2020 1390 10 10 0  6.28318530717959E+0000}%
   \special{sh 1}%
   \special{ar 2330 1370 10 10 0  6.28318530717959E+0000}%
   \special{sh 1}%
   \special{ar 2200 1520 10 10 0  6.28318530717959E+0000}%
   \special{sh 1}%
   \special{ar 2210 1340 10 10 0  6.28318530717959E+0000}%
   \special{sh 1}%
   \special{ar 1290 1720 10 10 0  6.28318530717959E+0000}%
   \special{sh 1}%
   \special{ar 1180 1780 10 10 0  6.28318530717959E+0000}%
   \special{sh 1}%
   \special{ar 680 1610 10 10 0  6.28318530717959E+0000}%
   \special{sh 1}%
   \special{ar 810 1640 10 10 0  6.28318530717959E+0000}%
   \special{sh 1}%
   \special{ar 970 1290 10 10 0  6.28318530717959E+0000}%
   \special{sh 1}%
   \special{ar 660 1400 10 10 0  6.28318530717959E+0000}%
   \special{sh 1}%
   \special{ar 520 1760 10 10 0  6.28318530717959E+0000}%
   \special{sh 1}%
   \special{ar 1810 2250 10 10 0  6.28318530717959E+0000}%
   \special{sh 1}%
   \special{ar 2000 2160 10 10 0  6.28318530717959E+0000}%
   \special{sh 1}%
   \special{ar 1600 2390 10 10 0  6.28318530717959E+0000}%
   \special{sh 1}%
   \special{ar 1750 2500 10 10 0  6.28318530717959E+0000}%
   \special{sh 1}%
   \special{ar 920 1090 10 10 0  6.28318530717959E+0000}%
   \special{sh 1}%
   \special{ar 2500 1010 10 10 0  6.28318530717959E+0000}%
   \special{sh 1}%
   \special{ar 2240 1030 10 10 0  6.28318530717959E+0000}%
   \special{sh 1}%
   \special{ar 2210 860 10 10 0  6.28318530717959E+0000}%
   \special{sh 1}%
   \special{ar 2610 1390 10 10 0  6.28318530717959E+0000}%
   \special{sh 1}%
   \special{ar 2720 2210 10 10 0  6.28318530717959E+0000}%
   \special{sh 1}%
   \special{ar 740 1770 10 10 0  6.28318530717959E+0000}%
   \special{sh 1}%
   \special{ar 1740 2030 10 10 0  6.28318530717959E+0000}%
   \special{sh 1}%
   \special{ar 1440 950 10 10 0  6.28318530717959E+0000}%
   \special{sh 1}%
   \special{ar 740 1050 10 10 0  6.28318530717959E+0000}%
   \special{sh 1}%
   \special{ar 520 1250 10 10 0  6.28318530717959E+0000}%
   \special{sh 1}%
   \special{ar 1160 1130 10 10 0  6.28318530717959E+0000}%
   %
   \special{pn 8}%
   \special{sh 1}%
   \special{ar 1000 1820 10 10 0  6.28318530717959E+0000}%
   \special{sh 1}%
   \special{ar 740 2080 10 10 0  6.28318530717959E+0000}%
   \special{sh 1}%
   \special{ar 880 2120 10 10 0  6.28318530717959E+0000}%
   \special{sh 1}%
   \special{ar 930 1900 10 10 0  6.28318530717959E+0000}%
   \special{sh 1}%
   \special{ar 1480 1860 10 10 0  6.28318530717959E+0000}%
   \special{sh 1}%
   \special{ar 1180 1920 10 10 0  6.28318530717959E+0000}%
   \special{sh 1}%
   \special{ar 1190 2080 10 10 0  6.28318530717959E+0000}%
   \special{sh 1}%
   \special{ar 880 1610 10 10 0  6.28318530717959E+0000}%
   \special{sh 1}%
   \special{ar 2040 1850 10 10 0  6.28318530717959E+0000}%
   \special{sh 1}%
   \special{ar 2000 2050 10 10 0  6.28318530717959E+0000}%
   \special{sh 1}%
   \special{ar 2350 2010 10 10 0  6.28318530717959E+0000}%
   \special{sh 1}%
   \special{ar 2520 1850 10 10 0  6.28318530717959E+0000}%
   \special{sh 1}%
   \special{ar 2240 1840 10 10 0  6.28318530717959E+0000}%
   \special{sh 1}%
   \special{ar 2320 1810 10 10 0  6.28318530717959E+0000}%
   \special{sh 1}%
   \special{ar 1780 1380 10 10 0  6.28318530717959E+0000}%
   \special{sh 1}%
   \special{ar 1660 1480 10 10 0  6.28318530717959E+0000}%
   \special{sh 1}%
   \special{ar 1890 1220 10 10 0  6.28318530717959E+0000}%
   \special{sh 1}%
   \special{ar 2080 1240 10 10 0  6.28318530717959E+0000}%
   \special{sh 1}%
   \special{ar 1910 1310 10 10 0  6.28318530717959E+0000}%
   \special{sh 1}%
   \special{ar 1930 900 10 10 0  6.28318530717959E+0000}%
   \special{sh 1}%
   \special{ar 2120 1040 10 10 0  6.28318530717959E+0000}%
   \special{sh 1}%
   \special{ar 1680 980 10 10 0  6.28318530717959E+0000}%
   \special{sh 1}%
   \special{ar 1440 1110 10 10 0  6.28318530717959E+0000}%
   \special{sh 1}%
   \special{ar 1760 1820 10 10 0  6.28318530717959E+0000}%
  \end{picture}%
 \end{center}
 \caption{D-instanton configuration associated with the one-loop
 processes in Fig.~\ref{fig:oneloop}} \label{fig:triangle}
\end{figure}

In order to illustrate how matrix perturbation theory works, let us
calculate the one loop contribution to two point function\footnote{no
sum is taken for indices $i$ or $j$; it is invariant under $U(1)^{N}$
gauge symmetry.}  $\langle X_{ij}^{\mu} X_{ji}^{\nu} \rangle$ using the
Feynman rules just derived.  Relevant fatgraphs are shown in
Fig.~\ref{fig:oneloop}.  In addition to the ``external'' D-instantons
$i$ and $j$, we need to incorporate an ``internal'' D-instanton $k$
as in Fig.~\ref{fig:triangle}.

Let us begin with the diagram (a). It can be seen as representing the
history of an open string; the string $ij$ splits into two pieces $ik$
and $kj$, and reconnect in the end.  Applying the Feynman rules, this
diagram contributes
\begin{eqnarray}
 \mbox{(a)} &=& -\frac{g^{2}}{x_{ij}^4}\sum_{k}\sum_{\kappa\lambda}
  \frac{1}{x_{ik}^{2} \, x_{jk}^{2}}\nonumber \\
 & & \times \{ \delta ^{\mu\kappa}(x_{ij}-x_{jk})^{\lambda}+
  \delta ^{\kappa\lambda}(x_{ik}-x_{kj})^{\mu}+
  \delta ^{\lambda\mu}(x_{ji}-x_{ik})^{\kappa} \} 
  \nonumber \\
 & & \times \{\delta ^{\kappa\nu}(x_{kj}-x_{ji})^{\lambda}
  +\delta ^{\nu\lambda}(x_{ji}-x_{ik})^{\kappa}+\delta ^{\lambda\kappa}
  (x_{ik}-x_{kj})^{\nu} \}.
  \label{bloop}
\end{eqnarray}
Here we denote by $\lambda$ and $\kappa$ the $SO(D)$ vector indices
associated to the upper and lower internal propagators respectively.
Similarly, from the diagrams (b) and (c), we have
\begin{equation}
 \mbox{(b)} + \mbox{(c)}
  = -\frac{g^{2}}{x_{ij}^4}\sum_{k} 
  \Bigl( \frac{1}{x_{ik}^2}
  + \frac{1}{x_{jk}^2} \Bigr)(1-D)\delta_{\mu\nu}.
\end{equation}
Ghosts contribute 
\begin{equation}
 \mbox{(d)}+\mbox{(e)}  = -\frac{2g^2}{x_{ij}^4}\sum_{k}
  \frac{x_{jk}^{\mu}x_{ik}^{\nu}}{x_{ik}^{2} \, x_{jk}^2}\; . \label{gloop}
\end{equation}
Finally fermion loop diagram gives
\begin{equation}
 \mbox{(f)}+\mbox{(g)} 
  = -\frac{d_{\Gamma}g^2}{2x_{ij}^4}\sum_{k}
  \frac{x_{jk}^{\mu}x_{ik}^{\nu}+x_{jk}^{\nu}x_{ik}^{\mu}
  -\delta ^{\mu\nu}x_{jk}\cdot x_{ik}}{x_{jk}^2 \, x_{ik}^2},
  \label{floop}
\end{equation}
where $d_{\Gamma}$ is the size of gamma matrices in $D$-dimensions.

 \section{Correspondence to quantum field theories}
\label{sec:toQFT}

From the sample calculation given in section \ref{sec:oneloop}, we
notice a strong similarity between the matrix perturbation theory and a
usual $d$-dimensional QFT.  Roughly speaking, relative brane position
$x_{ij}$ corresponds to a momentum $p$ whereas the sum over branes
$\sum_{k}$ looks like a loop integral $\int d^{d}p$.

In this section, we will make this analogy more precise.  In particular,
we illustrate how $d$-dimensional gauge theories can be recovered from
Matrix model, when D-instanton configuration has $d$-dimensional flat
directions.  In this context, the flat directions should be thought of
as momentum coordinates rather than spatial coordinates, contrary to
naive expectations.  This can be considered as an example of IR/UV
correspondence \cite{SW}.

Infinite sums or integrals will pose a delicate problem of large $N$
divergences.  We will postpone discussing this issues to section
\ref{sec:LargeN}.

  \subsection{D-instanton distribution as discretized momentum space}
\label{sec:DistAndMom}

We begin with recalling a general structure of $d$-dimensional gauge
theory amplitudes. Let $A^{\mu}(x)$ be $U(n)$ gauge fields represented
as $n \times n$ matrices and $A^{\mu}(p)$ their Fourier transform.
Hereafter the symbol ``$\tr$'' will denote the trace over $n \times n$
matrix indices. Any gauge invariant correlation function can be
decomposed as a sum of basic correlation functions of the form
\begin{equation}
 \langle  \, \tr
  A^{\mu_{1}}(p_{1})
  A^{\mu_{2}}(p_{2})
  \cdots
  A^{\mu_{k}}(p_{k})
  \,
  \rangle,
  \qquad 
  (p_{1}+\cdots+p_{k}=0)
  \label{YMAmp}
\end{equation}
which is invariant under the cyclic permutation of momenta and Lorenz
indices:
\begin{equation}
 \begin{split}
  & p_{1} \to p_{2} \to \cdots \to p_{k} \to p_{1},
  \\
  & \mu_{1} \to \mu_{2} \to \cdots \to \mu_{k} \to \mu_{1}.
 \end{split}
 \label{cyclicYM}
\end{equation}
In the amplitude (\ref{YMAmp}), the cyclic order of
$\{(p_{i},\mu_{i})\}_{i=1}^{k}$ has a definite meaning because $U(n)$
indices are implicitly contracted. It is easy to check that there is
one-to-one correspondence among the following three data:
\begin{enumerate}
 \item[\sf (A)] gauge invariant amplitude (\ref{YMAmp}),
 \item[\sf (B)] ordered set of momenta and Lorentz indices $\{
		(p_{1},\mu_{1}), (p_{2},\mu_{2}), \ldots, (p_{k},\mu_{k}) \}$
		modulo cyclic permutation (\ref{cyclicYM}),
 \item[\sf (C)] closed oriented path in $\R^{d}$ with edge labeling
		$\mu_{1},\cdots,\mu_{k}$
		(Fig.~\ref{fig:Polygon}~(a)).
\end{enumerate}
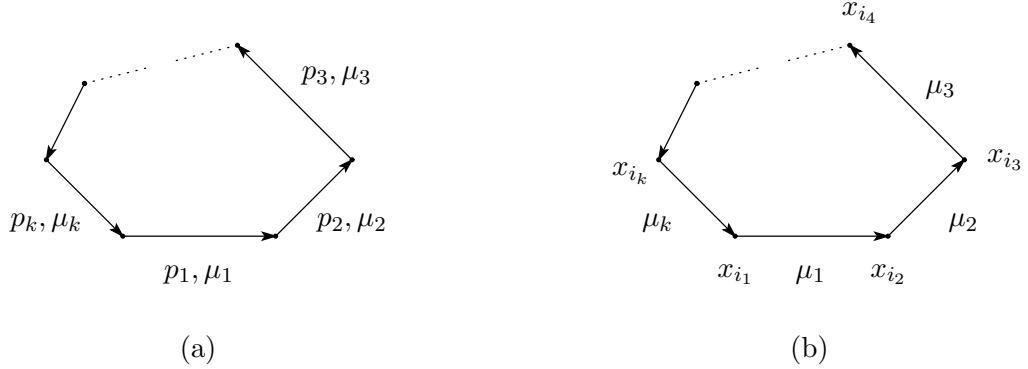
\begin{figure}[tbp]
 \begin{center}
  \unitlength 0.1in
  \begin{picture}(54.80,17.65)(1.30,-19.15)
   %
   \special{pn 8}%
   \special{sh 1}%
   \special{ar 1205 1405 10 10 0  6.28318530717959E+0000}%
   \special{sh 1}%
   \special{ar 2005 1405 10 10 0  6.28318530717959E+0000}%
   \special{sh 1}%
   \special{ar 2405 1005 10 10 0  6.28318530717959E+0000}%
   \special{sh 1}%
   \special{ar 1805 405 10 10 0  6.28318530717959E+0000}%
   \special{sh 1}%
   \special{ar 805 1005 10 10 0  6.28318530717959E+0000}%
   \special{sh 1}%
   \special{ar 1005 605 10 10 0  6.28318530717959E+0000}%
   \special{sh 1}%
   \special{ar 1005 605 10 10 0  6.28318530717959E+0000}%
   %
   \special{pn 8}%
   \special{pa 1205 1405}%
   \special{pa 2005 1405}%
   \special{fp}%
   \special{sh 1}%
   \special{pa 2005 1405}%
   \special{pa 1938 1385}%
   \special{pa 1952 1405}%
   \special{pa 1938 1425}%
   \special{pa 2005 1405}%
   \special{fp}%
   %
   \special{pn 8}%
   \special{pa 2005 1405}%
   \special{pa 2405 1005}%
   \special{fp}%
   \special{sh 1}%
   \special{pa 2405 1005}%
   \special{pa 2344 1038}%
   \special{pa 2367 1043}%
   \special{pa 2372 1066}%
   \special{pa 2405 1005}%
   \special{fp}%
   %
   \special{pn 8}%
   \special{pa 2405 1005}%
   \special{pa 1805 405}%
   \special{fp}%
   \special{sh 1}%
   \special{pa 1805 405}%
   \special{pa 1838 466}%
   \special{pa 1843 443}%
   \special{pa 1866 438}%
   \special{pa 1805 405}%
   \special{fp}%
   %
   \special{pn 8}%
   \special{pa 1005 605}%
   \special{pa 805 1005}%
   \special{fp}%
   \special{sh 1}%
   \special{pa 805 1005}%
   \special{pa 853 954}%
   \special{pa 829 957}%
   \special{pa 817 936}%
   \special{pa 805 1005}%
   \special{fp}%
   %
   \special{pn 8}%
   \special{pa 805 1005}%
   \special{pa 1205 1405}%
   \special{fp}%
   \special{sh 1}%
   \special{pa 1205 1405}%
   \special{pa 1172 1344}%
   \special{pa 1167 1367}%
   \special{pa 1144 1372}%
   \special{pa 1205 1405}%
   \special{fp}%
   %
   \special{pn 8}%
   \special{pa 1005 605}%
   \special{pa 1325 525}%
   \special{dt 0.045}%
   %
   \special{pn 8}%
   \special{pa 1805 405}%
   \special{pa 1485 485}%
   \special{dt 0.045}%
   \put(16.0500,-16.0500){\makebox(0,0){$p_{1},\mu_{1}$}}%
   \put(24.0500,-13.3500){\makebox(0,0){$p_{2},\mu_{2}$}}%
   \put(23.2500,-5.6500){\makebox(0,0){$p_{3},\mu_{3}$}}%
   \put(8.0500,-13.3500){\makebox(0,0){$p_{k},\mu_{k}$}}%
   %
   \special{pn 8}%
   \special{sh 1}%
   \special{ar 4410 1405 10 10 0  6.28318530717959E+0000}%
   \special{sh 1}%
   \special{ar 5210 1405 10 10 0  6.28318530717959E+0000}%
   \special{sh 1}%
   \special{ar 5610 1005 10 10 0  6.28318530717959E+0000}%
   \special{sh 1}%
   \special{ar 5010 405 10 10 0  6.28318530717959E+0000}%
   \special{sh 1}%
   \special{ar 4010 1005 10 10 0  6.28318530717959E+0000}%
   \special{sh 1}%
   \special{ar 4210 605 10 10 0  6.28318530717959E+0000}%
   \special{sh 1}%
   \special{ar 4210 605 10 10 0  6.28318530717959E+0000}%
   %
   \special{pn 8}%
   \special{pa 4410 1405}%
   \special{pa 5210 1405}%
   \special{fp}%
   \special{sh 1}%
   \special{pa 5210 1405}%
   \special{pa 5143 1385}%
   \special{pa 5157 1405}%
   \special{pa 5143 1425}%
   \special{pa 5210 1405}%
   \special{fp}%
   %
   \special{pn 8}%
   \special{pa 5210 1405}%
   \special{pa 5610 1005}%
   \special{fp}%
   \special{sh 1}%
   \special{pa 5610 1005}%
   \special{pa 5549 1038}%
   \special{pa 5572 1043}%
   \special{pa 5577 1066}%
   \special{pa 5610 1005}%
   \special{fp}%
   %
   \special{pn 8}%
   \special{pa 5610 1005}%
   \special{pa 5010 405}%
   \special{fp}%
   \special{sh 1}%
   \special{pa 5010 405}%
   \special{pa 5043 466}%
   \special{pa 5048 443}%
   \special{pa 5071 438}%
   \special{pa 5010 405}%
   \special{fp}%
   %
   \special{pn 8}%
   \special{pa 4210 605}%
   \special{pa 4010 1005}%
   \special{fp}%
   \special{sh 1}%
   \special{pa 4010 1005}%
   \special{pa 4058 954}%
   \special{pa 4034 957}%
   \special{pa 4022 936}%
   \special{pa 4010 1005}%
   \special{fp}%
   %
   \special{pn 8}%
   \special{pa 4010 1005}%
   \special{pa 4410 1405}%
   \special{fp}%
   \special{sh 1}%
   \special{pa 4410 1405}%
   \special{pa 4377 1344}%
   \special{pa 4372 1367}%
   \special{pa 4349 1372}%
   \special{pa 4410 1405}%
   \special{fp}%
   %
   \special{pn 8}%
   \special{pa 4210 605}%
   \special{pa 4530 525}%
   \special{dt 0.045}%
   %
   \special{pn 8}%
   \special{pa 5010 405}%
   \special{pa 4690 485}%
   \special{dt 0.045}%
   \put(44.1500,-16.0500){\makebox(0,0){$x_{i_1}$}}%
   \put(52.1500,-16.0500){\makebox(0,0){$x_{i_2}$}}%
   \put(50.6500,-2.3500){\makebox(0,0){$x_{i_4}$}}%
   \put(38.6500,-10.7500){\makebox(0,0){$x_{i_k}$}}%
   \put(58.2500,-10.0500){\makebox(0,0){$x_{i_3}$}}%
   \put(48.0500,-16.0500){\makebox(0,0){$\mu_{1}$}}%
   \put(56.0500,-13.3500){\makebox(0,0){$\mu_{2}$}}%
   \put(40.0500,-13.3500){\makebox(0,0){$\mu_{k}$}}%
   \put(54.8500,-6.3500){\makebox(0,0){$\mu_{3}$}}%
   \put(16.0000,-20.0000){\makebox(0,0){(a)}}%
   \put(48.0000,-20.0000){\makebox(0,0){(b)}}%
  \end{picture}%
 \end{center}
 \caption{Correspondence of a correlation function between gauge theory
 and Matrix model. (a): closed path formed by external momenta. (b):
 corresponding D-instanton configuration.} \label{fig:Polygon}
\end{figure}

We now come back to Matrix model correlation functions. In generic
background (\ref{waketa}) we still have unbroken $U(1)^{N}$ gauge
symmetry. Therefore, only gauge invariant ``Wilson loops'' such as
\begin{equation}
 \langle
  \tilde{X}^{\mu_{1}}_{i_{1}i_{2}}
  \tilde{X}^{\mu_{2}}_{i_{2}i_{3}}
  \cdots
  \tilde{X}^{\mu_{k}}_{i_{k}i_{1}}
  \rangle
  \label{MatrixAmp}
\end{equation}
can be nonzero. Actually, we can draw a corresponding loop as in
Fig.~\ref{fig:Polygon}~(b): two D-instantons located at $x_{i_{r}}$ and
$x_{i_{r+1}}$ are connected by a field (or an open string)
$\tilde{X}_{i_{r}i_{r+1}}^{\mu_{r}}$.  The amplitude (\ref{MatrixAmp})
is invariant under the cyclic permutation of D-instanton positions and
Lorenz indices:
\begin{equation}
 \begin{split}
  & i_{1} \to i_{2} \to \cdots \to i_{k} \to i_{1},
  \\
  & \mu_{1} \to \mu_{2} \to \cdots \to \mu_{k} \to \mu_{1}.
 \end{split}
 \label{cyclicM}
\end{equation}
The correspondence between the two figures Fig.~\ref{fig:Polygon}~(a)
and (b) is now obvious; the momentum $p_{r}$ in gauge theory is
identified with the D-instanton separation $x_{i_{r+1}}-x_{i_{r}}$. In
this way, we can add two new entries to the previous list of one-to-one
correspondence:
\begin{enumerate}
 \item[\sf (D)] D-instanton correlation functions (\ref{MatrixAmp}),

 \item[\sf (E)] loop passing through $k$ D-instantons in $\R^{d}$ with
		labeled edges $\mu_{1},\ldots,\mu_{k}$ \newline
		(Fig.~\ref{fig:Polygon}~(b)).
\end{enumerate}
Clearly, arbitrary sequence of $d$-dimensional momenta $(p_{1},\ldots,
p_{k})$ can be realized as a loop in D-instanton configuration space,
provided D-instantons are densely distributed over $d$-dimensional
Euclidean space $\R^{d}$. Conversely, if we start from finite $N$ matrix
theory, only discrete momentum points are available on gauge theory
side. Of course, the error becomes smaller as the number of D-instantons
is increased. Guided by these observation, we propose to \emph{identify
a D-instanton distribution of Matrix theory as a discretization of
momentum space seen by a Yang-Mills theory.}

Precisely speaking, the ``momentum path'' (Fig.~\ref{fig:Polygon}~(a))
can determine the ``D-instanton path'' (Fig.~\ref{fig:Polygon}~(b)) only
up to overall translation, $x_{i}\to x_{i}+c$. This ambiguity can be
resolved by fixing, say, their center of mass at the origin.

So far we neglected the problem how non-Abelian $U(n)$ gauge symmetry
can be recovered from D-instanton picture.  This will be discussed in
section \ref{sec:NonAbel}.

In sum, we have argued that when the background D-instantons are
continuously distributed along $\R^{d}$, there is a one-to-one
correspondence between the gauge theory amplitude (\ref{YMAmp}) and
(\ref{MatrixAmp}).

  \subsection{Correspondence of Feynman diagrams}
\label{sec:FeynmanCorr}

In perturbation theory, both amplitudes (\ref{YMAmp}) and
(\ref{MatrixAmp}) are expressed as a sum over fatgraphs with fixed
external lines. We now want to show that two computations, one as a
Yang-Mills theory and the other as a Matrix theory, actually coincide
for every fatgraph. To do this, we need to check the correspondence at
the level of propagators and vertices.

Consider a fatgraph $\Gamma$ made of several propagators and
vertices. Recall that the graph form an oriented Riemann surface with
boundaries.  Thus for a given propagator with an orientation, it is
meaningful to talk about its ``left-'' and ``right-'' edges.  

Pick up a propagator and let $i$ and $j$ be its labels on left- and
right-edges, repectively. In Matrix picture, the propgator represents a
fluctuation $X_{ij}^{\mu}$ connecting two D-instantons $i$ and $j$.  As
in section \ref{sec:DistAndMom}, we identify the relative separation of
D-instantons
\begin{equation}
 x_{ij}=x_{i}-x_{j}
\end{equation}
with the momentum carried by the propagator in a corresponding QFT.

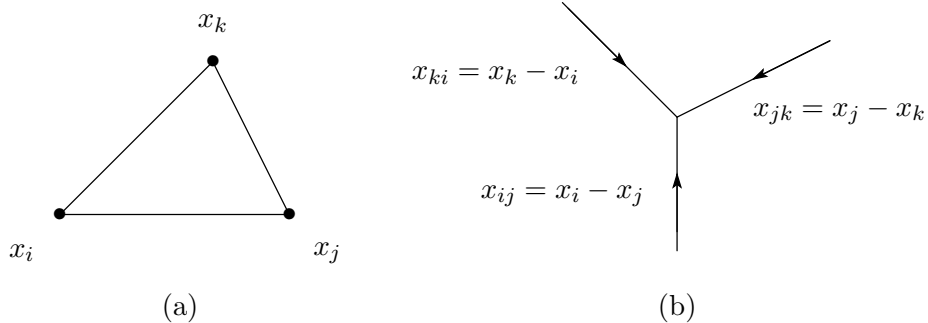
\begin{figure}[btp]
 \begin{center}
  \unitlength 0.1in
  \begin{picture}(45.45,15.15)(2.85,-24.05)
   %
   \special{pn 8}%
   \special{pa 800 2000}%
   \special{pa 2000 2000}%
   \special{pa 1600 1200}%
   \special{pa 1600 1200}%
   \special{pa 800 2000}%
   \special{fp}%
   \put(16.0000,-12.0000){\makebox(0,0){$\bullet$}}%
   \put(20.0000,-20.0000){\makebox(0,0){$\bullet$}}%
   \put(8.0000,-20.0000){\makebox(0,0){$\bullet$}}%
   %
   \special{pn 8}%
   \special{pa 4030 1490}%
   \special{pa 3430 890}%
   \special{fp}%
   %
   \special{pn 8}%
   \special{pa 4030 1490}%
   \special{pa 4030 2090}%
   \special{fp}%
   %
   \special{pn 8}%
   \special{pa 4030 1490}%
   \special{pa 4830 1090}%
   \special{fp}%
   %
   \special{pn 8}%
   \special{pa 4830 1090}%
   \special{pa 4430 1290}%
   \special{fp}%
   \special{sh 1}%
   \special{pa 4430 1290}%
   \special{pa 4499 1278}%
   \special{pa 4478 1266}%
   \special{pa 4481 1242}%
   \special{pa 4430 1290}%
   \special{fp}%
   %
   \special{pn 8}%
   \special{pa 3460 920}%
   \special{pa 3760 1220}%
   \special{fp}%
   \special{sh 1}%
   \special{pa 3760 1220}%
   \special{pa 3727 1159}%
   \special{pa 3722 1182}%
   \special{pa 3699 1187}%
   \special{pa 3760 1220}%
   \special{fp}%
   %
   \special{pn 8}%
   \special{pa 4030 2190}%
   \special{pa 4030 1790}%
   \special{fp}%
   \special{sh 1}%
   \special{pa 4030 1790}%
   \special{pa 4010 1857}%
   \special{pa 4030 1843}%
   \special{pa 4050 1857}%
   \special{pa 4030 1790}%
   \special{fp}%
   \put(6.0000,-22.0000){\makebox(0,0){$x_{i}$}}%
   \put(22.0000,-22.0000){\makebox(0,0){$x_{j}$}}%
   \put(16.0000,-10.0000){\makebox(0,0){$x_{k}$}}%
   \put(34.3000,-18.9000){\makebox(0,0){$x_{ij}=x_{i}-x_{j}$}}%
   \put(48.8000,-14.6000){\makebox(0,0){$x_{jk}=x_{j}-x_{k}$}}%
   \put(30.8000,-12.6000){\makebox(0,0){$x_{ki}=x_{k}-x_{i}$}}%
   \put(14.3000,-24.9000){\makebox(0,0){(a)}}%
   \put(40.3000,-24.9000){\makebox(0,0){(b)}}%
  \end{picture}%
 \end{center}
 \caption{(a) D-instanton configuration and (b) corresponding momentum
 flow} \label{fig:DAndMom}
\end{figure}
This identification would be incocnsistent if there was an interaction
vertex which breaks their conservation law. In fact, any interaction
vertex shown in Fig.~\ref{fig:vertex} has following properties: (i)
edges of the propagators are glued together in a definite cyclic order
around the vertex so that successive edges share a common D-instanton
label, (ii) all incoming momenta are given as the differences of
$x_{i}$'s associated to successive boundaries.  These two facts
guarantee the sum of the incoming momenta is automatically
zero.\footnote{The momentum conservation is a direct consequence of
$U(1)^{N}$ gauge invariance. It is, however, nontrivial whether one can
represent the momenta as differences successively.}  For example,
consider an interaction vertex among three D-instantons depicted in
Fig.~\ref{fig:DAndMom}~(a). Since the incoming momenta are defined as
Fig.~\ref{fig:DAndMom}~(b), their sum $x_{ij}+x_{jk}+x_{ki}$ vanishes.
It is now straightforward to check that Feynman rules given in
Figs. \ref{fig:propagator} and \ref{fig:vertex} exactly coincide with
those of a usual Yang-Mills theory.

  \subsection{From sums to integrals}
\label{sec:Sum2Int}

In an ordinary perturbation method, we need to integrate over
interaction positions in $d$-dimensional spacetime, which results in
loop integrals. In the matrix perturbation theory, we need to sum over
intermediate D-instanton positions. The correspondence between these two
implies the equivalence of perturbation theory, because we have just
seen that each fatgraph has the same factors both in Matrix and QFT
pictures.

As discussed in section \ref{sec:DistAndMom}, D-instanton configuration
along flat $\R^{d}$ can reproduce $d$-dimensional momentum space.  More
concretely, let us assume D-instantons fill uniformly a $d$-dimensional
hyperplane
\begin{equation}
 H := \left\{(x^{1},\ldots,x^{d},0,\ldots,0)\in \R^{D}\right\}
  \simeq \R^{d}
  \label{Hyperdef}
\end{equation}
with density $\rho$ (i.e. there are $\rho$ D-instantons per unit
$d$-dimensional volume). Then, we can replace the sum over brane
positions $\sum_{k}$ by the loop integral $\rho \int d^{d}p$.

For example, in the case of the two point function $\langle
X_{ij}^{\mu}X_{ji}^{\nu}\rangle$ of section \ref{sec:oneloop}, we can
choose the point $x_{i}$ as the origin of the momentum space. Then the
translation dictionary reads
\begin{equation}
 \begin{array}{lll} \mbox{Matrix} & & \mbox{QFT} \\ 
  x_{ij} & \Longrightarrow & q \quad (\mbox{external momentum}) \\
  x_{jk} & \Longrightarrow & p \quad (\mbox{loop momentum}) \\
  x_{ik} & \Longrightarrow & q+p  \\
  \displaystyle\sum_{k}\cdots & \Longrightarrow & \rho
   \displaystyle \int d^{d}p \cdots
 \end{array}
 \label{Sum2Int}
\end{equation}
Then eqs. (\ref{bloop}), (\ref{gloop}), (\ref{floop}) can be written as
\begin{eqnarray}
 \mbox{(a)} &=& -\frac{ \rho g^{2}}
  {q^4}\int d^{d}p \;  
  \frac{1}{(q+p)^{2}\cdot p^{2}}\nonumber \\
 & & \times \{ \delta ^{\mu\kappa}(q-p)^{\lambda}+
  \delta ^{\kappa\lambda}(2q+p)^{\mu}+
  \delta ^{\lambda\mu}(-q-2p)^{\kappa} \} \nonumber \\
 & & \times \{\delta ^{\kappa\nu}(p-q)^{\lambda}
  +\delta ^{\nu\lambda}(p+2q)^{\kappa}+\delta ^{\lambda\kappa}(-2p-q)^{\nu} \}  
  \nonumber  \\
 \nonumber  \\
 \mbox{(b)+(c)} &=& -\frac{2 \rho g^2}{q^4}
  \int d^{d}p \; \left(\frac{1}{(q+p)^2}+\frac{1}{p^2} \right)(1-d)\delta_{\mu\nu}
  \nonumber    \\
 \nonumber  \\
 \mbox{(d)+(e)} &=& -\frac{2 \rho g^2}{q^4}
  \int d^{d}p \;  \frac{p^{\mu}(q+p)^{\nu}}{(q+p)^{2}p^2} 
  \nonumber    \\
 \nonumber  \\
 \mbox{(f)+(g)} &=& -\frac{d_{\Gamma} \rho g^2}{2q^4}
  \int d^{d}p \; \frac{p^{\mu}(q+p)^{\nu}+
  p^{\nu}(q+p)^{\mu}-\delta ^{\mu\nu} p \cdot (q+p)}{p^2(q+p)^2}
\end{eqnarray}
which look more familiar as those in standard QFT textbooks.

  \subsection{Non-Abelian gauge symmetry in Matrix model}
\label{sec:NonAbel}

Now we come back to the problem how we can incorporate the non-Abelian
gauge symmetry of $d$-dimensional SYM theory starting from Matrix
models. Actually, without this non-Abelian structure, we cannot explain
why the cyclic order of momenta is important in the amplitudes
(\ref{YMAmp}).

To achieve this, we need to consider the coincident D-branes as in
\cite{Witten}.  Suppose we want to realize $U(n)$ gauge symmetry. We
need to put $n$ D-instantons at the same point in $\R^{D}$.  Hereafter
the word ``cluster'' will be used to designate the $n$ coincident
D-instantons.  The $N$ D-instantons are thus grouped into $M \equiv N/n$
clusters.

We choose a background in which cluster $r$ is located at
$x_{r}=(x_{r}^{1},\ldots,x_{r}^{d},0,\ldots,0) \in \R^{D}$: \footnote{We
use $r,s,\ldots \in \{1,\ldots,M\}$ for cluster indices.  $i,j,\ldots
\in \{1,\ldots,N\}$ are reserved for D-instanton labels.}
\begin{equation}
 \begin{array}{rll}
  \bar{X}^{\mu} &=
   \left( 
    \begin{array}{c|c|c|c}
     \strut 
      x_{1}^{\mu}{\bf 1} & {\bf 0} &\cdots  & {\bf 0} \\
     \hline
      \strut 
      {\bf 0} & x_{r}^{\mu}{\bf 1} & \ddots & \vdots   \\
     \hline
      \strut 
      \vdots & \ddots & \ddots & {\bf 0} \\ 
     \hline
      \strut 
      {\bf 0} & \cdots & {\bf 0} & x_{M}^{\mu}{\bf 1}
    \end{array} 
    \right) 
    &(\mu=1,\ldots,d),
    \\
  \bar{X}^{m}
   &
   =0 
   &
   (m=d+1,\ldots,D).
 \end{array}
 \label{YMbackgroundX}
\end{equation}
Here ${\bf 1}$ and ${\bf 0}$ denote unit and zero matrix of size $n$,
respectively.

In this background, $U(N)$ gauge symmetry is broken down to $U(n)^{M}$
generated by
\begin{equation}
 U =
  \left( 
   \begin{array}{c|c|c|c}
    \strut 
     U_{11} & {\bf 0} & \cdots & {\bf 0}  \\
    \hline
     \strut 
     {\bf 0} & U_{rr} & \ddots & \vdots \\
    \hline
     \strut 
     \vdots & \ddots & \ddots & {\bf 0} \\ 
    \hline
     \strut 
     {\bf 0} &\cdots & {\bf 0} & U_{MM} 
   \end{array} 
   \right).
\end{equation}

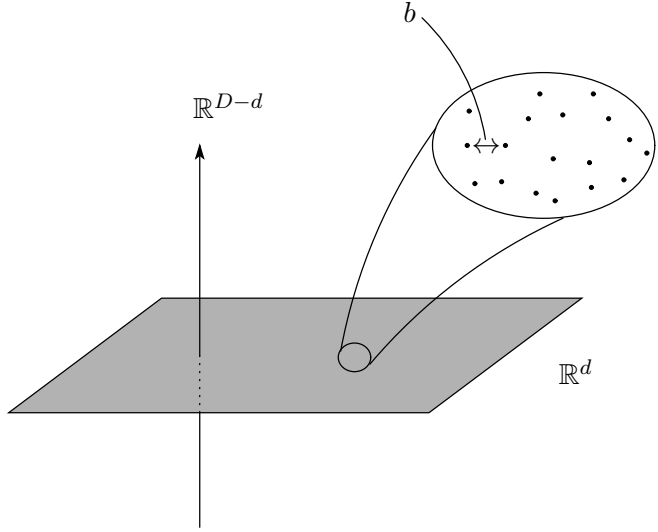
\begin{figure}[tbp]
 \begin{center}
  \unitlength 0.1in
  \begin{picture}(33.83,27.85)(6.00,-30.00)
   %
   \special{pn 8}%
   \special{sh 0.300}%
   \special{pa 600 2400}%
   \special{pa 1400 1800}%
   \special{pa 3600 1800}%
   \special{pa 2800 2400}%
   \special{pa 2800 2400}%
   \special{pa 2800 2400}%
   \special{pa 600 2400}%
   \special{fp}%
   %
   \special{pn 8}%
   \special{ar 2410 2110 85 75  0.0000000 6.2831853}%
   %
   \special{pn 8}%
   \special{ar 3400 1000 583 383  0.0000000 6.2831853}%
   %
   \special{pn 8}%
   \special{ar 5184 2600 2894 2894  3.3231644 3.7658508}%
   %
   \special{pn 8}%
   \special{ar 4699 4014 2893 2893  3.8436650 4.2859041}%
   %
   \special{pn 8}%
   \special{sh 1}%
   \special{ar 3010 820 10 10 0  6.28318530717959E+0000}%
   \special{sh 1}%
   \special{ar 3040 1200 10 10 0  6.28318530717959E+0000}%
   \special{sh 1}%
   \special{ar 3180 1190 10 10 0  6.28318530717959E+0000}%
   \put(34.8000,-22.5000){\makebox(0,0)[lb]{$\R ^d$}}%
   \put(17.5000,-7.9000){\makebox(0,0){$\R^{D-d}$}}%
   %
   \special{pn 8}%
   \special{sh 1}%
   \special{ar 3000 1000 10 10 0  6.28318530717959E+0000}%
   \special{sh 1}%
   \special{ar 3200 1000 10 10 0  6.28318530717959E+0000}%
   \special{sh 1}%
   \special{ar 3360 1250 10 10 0  6.28318530717959E+0000}%
   \special{sh 1}%
   \special{ar 3640 1090 10 10 0  6.28318530717959E+0000}%
   \special{sh 1}%
   \special{ar 3450 1070 10 10 0  6.28318530717959E+0000}%
   \special{sh 1}%
   \special{ar 3460 1290 10 10 0  6.28318530717959E+0000}%
   \special{sh 1}%
   \special{ar 3820 1180 10 10 0  6.28318530717959E+0000}%
   \special{sh 1}%
   \special{ar 3650 1220 10 10 0  6.28318530717959E+0000}%
   \special{sh 1}%
   \special{ar 3660 730 10 10 0  6.28318530717959E+0000}%
   \special{sh 1}%
   \special{ar 3500 840 10 10 0  6.28318530717959E+0000}%
   \special{sh 1}%
   \special{ar 3850 970 10 10 0  6.28318530717959E+0000}%
   \special{sh 1}%
   \special{ar 3740 860 10 10 0  6.28318530717959E+0000}%
   \special{sh 1}%
   \special{ar 3380 730 10 10 0  6.28318530717959E+0000}%
   \special{sh 1}%
   \special{ar 3320 860 10 10 0  6.28318530717959E+0000}%
   \special{sh 1}%
   \special{ar 3940 1040 10 10 0  6.28318530717959E+0000}%
   \put(31.0000,-10.1000){\makebox(0,0){$\leftrightarrow$}}%
   \put(27.0000,-3.0000){\makebox(0,0){$b$}}%
   %
   \special{pn 8}%
   \special{ar 1970 1150 1140 1140  5.4819926 6.1227971}%
   %
   \special{pn 8}%
   \special{pa 1600 2100}%
   \special{pa 1600 1000}%
   \special{fp}%
   \special{sh 1}%
   \special{pa 1600 1000}%
   \special{pa 1580 1067}%
   \special{pa 1600 1053}%
   \special{pa 1620 1067}%
   \special{pa 1600 1000}%
   \special{fp}%
   %
   \special{pn 8}%
   \special{pa 1600 2100}%
   \special{pa 1600 2400}%
   \special{dt 0.045}%
   %
   \special{pn 8}%
   \special{pa 1600 2400}%
   \special{pa 1600 3000}%
   \special{fp}%
  \end{picture}%
 \end{center}
 \caption{D-instanton configuration corresponding to $d$-dimensional
 SYM. Each cluster is denoted by a blob.  The clusters are distributed
 along $\R ^{d}$ with average spacing $b$.}  
 \label{fig:finiteN}
\end{figure}

As for fluctuations, it is useful to divide $N\times N$ matrix into the
blocks of size $n \times n$ as
\begin{equation}
 \begin{split}
  \tilde{X}^{\mu} & =
  \left( 
  \begin{array}{c|c|c|c}
   \strut 
    \tilde{X}_{11}^{\mu} & \cdots & \cdots  & \tilde{X}_{1M}^{\mu} \\
   \hline
    \strut 
    \vdots & \ddots & \tilde{X}_{rs}^{\mu} & \vdots \\
   \hline
    \strut 
    \vdots & \ddots & \ddots & \vdots \\ 
   \hline
    \strut 
    \tilde{X}_{M1}^{\mu} & \cdots & \cdots & \tilde{X}_{MM}^{\mu} 
  \end{array} 
  \right)
  \qquad
  (\mu=1,\ldots,d),
  \\ 
  \tilde{X}^{m} & =
  \rule{0pt}{19mm}
  \left( 
  \begin{array}{c|c|c|c}
   \strut 
    \tilde{X}_{11}^{m} & \cdots & \cdots  & \tilde{X}_{1M}^{m} \\
   \hline
    \strut 
    \vdots & \ddots & \tilde{X}_{rs}^{m} & \vdots \\
   \hline
    \strut 
    \vdots & \ddots & \ddots & \vdots \\ 
   \hline
    \strut 
    \tilde{X}_{M1}^{m} & \cdots & \cdots & \tilde{X}_{MM}^{m} 
  \end{array} 
  \right)
  \qquad
  (m=d+1,\ldots,D),
  \\
  \psi & =
  \rule{0pt}{19mm}
  \left( 
  \begin{array}{c|c|c|c}
   \strut 
    \psi_{11} & \cdots & \cdots & \psi_{1M}  \\
   \hline
    \strut 
    \vdots & \ddots & \psi_{rs} & \vdots \\
   \hline
    \strut 
    \vdots & \ddots & \ddots & \vdots \\ 
   \hline
    \strut 
    \psi_{M1} &\cdots &\cdots & \psi_{MM} 
  \end{array} 
  \right).
  \label{YMfluctuation}
 \end{split}
\end{equation}
Each $n \times n$ block $\tilde{X}_{rs}^{\mu}$, $\psi_{rs}$ transforms
as a bi-fundamental representation of $U(n)_{r}\times U(n)_{s}$ subgroup
of $U(n)^{M}$.  As we will see shortly (cf. eq. (\ref{SYMsubs}) below),
the fluctuations $\tilde{X}_{rs}^{\mu} \; (\mu=1,\ldots,d)$ tangent to
$H$ will be identified with the $U(n)$ gauge field whereas those in the
transverse direction $\tilde{X}_{rs}^{m} \; (m=d+1,\ldots,D)$ will play
the role of Higgs fields. Similarly, $D$-dimensional spinor $\psi_{rs}$
will be identified with their super partners.

In this notation, $U(n)^M$ invariant correlation functions are something
like
\begin{equation}
 \langle \, \tr X_{r_{1}r_{2}}^{\mu_{1}} X_{r_{2}r_{3}}^{\mu_{2}} \cdots
  X_{r_{k}r_{1}}^{\mu_{k}} \, \, \rangle.
\end{equation}
The correspondence given in sections \ref{sec:DistAndMom},
\ref{sec:FeynmanCorr} and \ref{sec:Sum2Int} is true to this non-Abelian
case, provided ``D-instantons'' are now replaced by ``clusters.''

  \subsection{$d$-dimensional super Yang-Mills action from Matrix model}
\label{sec:ddimSYM}

So far, we have studied how a gauge theory is embedded into Matrix
theory, through the correspondence of correlation functions.

To complete our analysis and to extract further intuition, it will be
useful to rewrite the original Matrix model action into that of QFT:
$d$-dimensional reduction of SYM in $D$ dimensions.

In the notation given in (\ref{YMbackgroundX}) and
(\ref{YMfluctuation}), the action (\ref{saisyo}) reads
\begin{eqnarray}
 S &=& -\frac{1}{g^2}\sum_{r,s} \tr
  \Bigl\{\frac{1}{4}
   (x_{rs}^{\mu}\tilde{X}_{rs}^{\nu}-x_{rs}^{\nu}\tilde{X}_{rs}^{\mu}
   +[\tilde{X}^{\mu},\tilde{X}^{\nu}]_{rs})
   (x_{sr}^{\mu}\tilde{X}_{sr}^{\nu}-x_{sr}^{\nu}\tilde{X}_{sr}^{\mu}
   +[\tilde{X}^{\mu},\tilde{X}^{\nu}]_{sr}) \nonumber\\
 && \qquad\qquad\quad  
  +\frac{1}{2}(x_{rs}^{\mu}\tilde{X}_{rs}^{m}
  +[\tilde{X}^{\mu},\tilde{X}^{m}]_{rs})
  (x_{sr}^{\mu}\tilde{X}_{sr}^{m}
  +[\tilde{X}^{\mu},\tilde{X}^{m}]_{sr})\nonumber\\
 && \qquad\qquad\quad    
  +\frac{1}{4}[\tilde{X}^{m},\tilde{X}^{n}]_{rs}
  [\tilde{X}^{m},\tilde{X}^{n}]_{sr}
  \label{MatrixActionYM}\\
 && \qquad\qquad\quad    
  +\frac{1}{2}\bar{\psi}_{sr}\Gamma^{\mu}
  (x_{rs}^{\mu}\psi_{rs}+[\tilde{X}^{\mu}, \psi]_{rs})\nonumber\\
 && \qquad\qquad\quad    
  +\frac{1}{2}\bar{\psi}_{sr}\Gamma^{m}[\tilde{X}^{m}, \psi]_{rs}
  \Bigr\}   
  \nonumber
\end{eqnarray} 
where the trace ``$\tr$'' is taken over $n \times n$ matrix indices.

Let us approximate the sum by the $d$-dimensional integral, which looks
more like a QFT. As in section \ref{sec:NonAbel}, we assume the clusters
$\{ x_{r} \}$ are uniformly distributed on $d$-dimensional hyperplane
$H$ with a constant cluster density $\rho '$.  Since each cluster
consists of $n$ D-instantons, $\rho'$ is related to the D-instanton
density $\rho$ as $n \rho'=\rho$.

Renaming the $n\times n$ matrix valued fields as
\begin{equation}
 \begin{array}{ccl}
  \tilde{X}_{rs}^{\mu} 
   &\Longrightarrow 
   &\tilde{X}^{\mu}(p)\equiv A^{\mu}(p)
   \\
  \tilde{X}_{rs}^{\mu} 
   &\Longrightarrow 
   &\tilde{X}^{m}(p)\equiv \Phi^{m}(p)
   \\
  \psi_{rs} &\Longrightarrow & \psi(p)
 \end{array}
 \label{SYMsubs}
\end{equation}
and the continuum approximation similar to (\ref{Sum2Int}),
\begin{equation}
 \begin{split}
  x_{rs}^{\mu} & \Longrightarrow p^{\mu}
  \\
  \sum_{s} \cdots 
  & \Longrightarrow \rho' \int d^{d}p \tr \cdots,
 \end{split}
\end{equation}
the action (\ref{MatrixActionYM}) reads
\begin{equation}
 \begin{split}
  S= 
  & 
  -\frac{(\sum_{r}1)}{g^2}\rho'  \int d^{d}p \; 
  \\
  & 
  \tr \Bigl\{
  \frac{1}{4}
  (p^{\mu}\tilde{X}^{\nu}(p)-p^{\nu}\tilde{X}^{\mu}(p)
  +[\tilde{X}^{\mu},\tilde{X}^{\nu}](p))
  (-p^{\mu}\tilde{X}^{\nu}(-p)+p^{\nu}\tilde{X}^{\mu}(-p)
  +[\tilde{X}^{\mu},\tilde{X}^{\nu}](-p)) \\
  &  +\frac{1}{2}(p^{\mu}\tilde{X}^{m}(p)
  +[\tilde{X}^{\mu},\tilde{X}^{m}](p))
  (-p^{\mu}\tilde{X}^{m}(-p)
  +[\tilde{X}^{\mu},\tilde{X}^{m}](-p))\\
  &  +\frac{1}{4}[\tilde{X}^{m},\tilde{X}^{n}](p)
  [\tilde{X}^{m},\tilde{X}^{n}](-p)\\
  &  +\frac{1}{2}\bar{\psi}(-p)\Gamma^{\mu}
  (p^{\mu}\psi(p)+[\tilde{X}^{\mu}, \psi](p))\\
  &  +\frac{1}{2}\bar{\psi}(-p)\Gamma^{m}[\tilde{X}^{m}, \psi](p)
  \Bigr\} .
  \label{SinMomentum}
 \end{split}
\end{equation}
This is a momentum space representation of the super Yang-Mills action
in $d$-dimensions.

Using formula like
\begin{eqnarray}
 x^{\mu}(p) \tilde{X}^{\nu}(p) &=& -i \int d^{d}x (\partial^{\mu}
  A^{\nu}(x)) e^{-ip\cdot x} \nonumber
  \\{}
  [\tilde{X}^{\mu},\tilde{X}^{\nu}](p) &=& 
  \rho' (2\pi)^{d} \int d^{d}x[A^{\mu}(x),A^{\nu}(x)]
  e^{-ip\cdot x},
\end{eqnarray}
inverse Fourier transform of (\ref{SinMomentum}) gives
\begin{eqnarray}
 \lefteqn{
  S = \frac{\left(\sum_{r}1\right)}{g^{2}}
  \rho'
  (2\pi)^{d} \int d^{d}x}
  \nonumber \\
 &&
  \quad \times \tr \biggl[
  \frac{1}{4}
  \left\{
   ( \partial^{\mu}A^{\nu}(x) - \partial^{\nu}A^{\mu}(x) )
   + i\rho'
   (2\pi)^{d} [A^{\mu}(x), A^{\nu}(x)]
\right\}^{2}
 \nonumber \\
 && \qquad +
  \frac{1}{2}
  \left\{
   ( \partial^{\mu}\Phi^{m}(x)
   + i\rho'
   (2\pi)^{d} [A^{\mu}(x), \Phi^{m}(x)]
\right\}^{2}
 \nonumber  \\
 && \qquad -
  \frac{1}{4}
  \left\{
   \rho'
   (2\pi)^{d} [\Phi^{m}(x), \Phi^{n}(x)]
\right\}^{2}
 \label{SYMaction}
 \\
 && \qquad +
  \frac{i}{2}
  \left\{ \bar{\psi}(x) \Gamma^{\mu}
   ( \partial^{\mu}\psi(x)
   + i\rho'
   (2\pi)^{d} [A^{\mu}(x), \psi(x)]
\right\} 
 \nonumber  \\
 && \qquad -
  \frac{1}{2}
  \left\{ \bar{\psi}(x) \Gamma^{m}
   \rho'
   (2\pi)^{d} [\Phi^{m}(x), \psi(x)]
\right\}
 \biggr].
 \nonumber 
\end{eqnarray}
Rescaling the fields as
\begin{equation}
 \begin{array}{rcl}
  \rho'(2\pi)^{d} A^{\mu}(x) &\rightarrow & A^{\mu}(x)\\
  \rho'(2\pi)^{d} \Phi^{m}(x) &\rightarrow & \Phi^{m}(x)\\
  \rho'(2\pi)^{d} \psi(x) &\rightarrow & \psi(x)\\
 \end{array}
\end{equation}
and defining the $d$-dimensional coupling constant by
\begin{equation}
 g_{d}^{2} \equiv (2\pi)^{d} \rho' g^{2}, 
  \label{gddef}
\end{equation}
we are finally lead to a familiar form of $d$-dimensional SYM coupled
with adjoint matters:\footnote{We neglected the factor
$(\sum_{r}1)$. This reflects the fact that overall shift of D-instanton
configuration results in the same momentum configuration in SYM picture. 
To make the mapping one-to-one, we need to specify the origin of
momentum space as discussed toward the end of section
\ref{sec:DistAndMom}. (Admittedly, the replacement (\ref{Sum2Int}) is
somewhat misleading.) Other possibility is to introduce an additional
gauge symmetry to constrain the off-diagonal components as
$X_{r,s}=X_{r+k,s+k}\; (\forall k \in \Z)$. This gauge symmetry kills
the ambiguity of overall shift, but the clusters need to be arranged
periodically on a lattice.  The latter method is equivalent to the
$S^{1}$ compactification proposed by W. Taylor \cite{Taylor}.}
\begin{eqnarray}
 \lefteqn{S_{d}=\int d^{d}x \frac{1}{g_{d}^{2}}
  \tr \biggl\{
  \frac{1}{4}(F^{\mu\nu})^{2} 
  + \frac{1}{2}(D^{\mu}\Phi^{m})^{2}
  -\frac{1}{4} [\Phi^{m},\Phi^{n}]^{2}}
  \nonumber\\
 && \qquad 
  +\frac{i}{2}\bar{\psi}\Gamma^{\mu}(D^{\mu}\psi) 
  -\frac{1}{2}\bar{\psi}\Gamma^{m}[\Phi^{m},\psi]
  \biggr\}
  \label{dsym}
\end{eqnarray}
Here, the standard covariant derivative for the adjoint matter $
D^{\mu}\Phi^{m} = \partial^{\mu}\Phi^{m} + i [A^{\mu}, \Phi^{m}]$ and
the field strength $F^{\mu\nu} = \partial^{\mu}A^{\nu} -
\partial^{\nu}A^{\mu} + i [A^{\mu}, A^{\nu}]$ is used.

Our procedure can be schematically summarized as
\begin{equation}
 \sum_{i} \cdots 
  \Longleftrightarrow \sum_{s}\tr \cdots 
  \Longleftrightarrow \rho' \int d^{d}p \tr \cdots .
  \label{SubsSummary}
\end{equation}

Now we make a few comments on the hermiticity of the fields.  If we
naively interpreted the brane configuration as the discretization in the
{\em spatial} coordinates (\`{a} la lattice gauge theory) rather than
the momentum coordinates, then an $n\times n$ block $X_{rs}^{\mu}$ must
be interpreted as a $U(n)$ connection (or parallel transport) matrix
connecting two points $x_{r}$ and $x_{s}$. Then it must be
anti-hermitian (or unitary) matrix. But, since $X_{rs}^{\mu}$ is just a
part of a much larger $N\times N$ hermitian matrix, there is no a priori
reason why it should take such special forms. It might be possible to
devise a mechanism to put such a constraint, it would lead to additional
complication.  Actually, the hermiticity as a $N\times N$ matrix leads
to $(\tilde{X}_{rs}^{\mu})^{\dagger}=\tilde{X}_{sr}^{\mu}$.  This is
consistent with our identification $X_{rs}^{\mu}$ with $A^{\mu}(p)$,
because the hermiticity of the gauge field $A^{\mu}(x)=\int
(\frac{dp}{2\pi})^{d} (A^{\mu}(p)e^{ip\cdot x}+ h.c. )$ leads to the
condition $(A^{\mu}(p))^{\dagger}=A^{\mu}(-p)$.

Note also that the diagonal parts $X_{ii}^{\mu}$ of $N \times N$ matrix
field correspond to the zero-momentum, Cartan component of $U(N)$
gauge field.

 \section{Divergences in large $N$ matrix theory}
\label{sec:LargeN}

Although the Matrix model is defined for arbitrary value of $N$, 
interesting physics is believed to emerge from the large $N$ limit.  But
there seems to be no general argument or rule concerning what kind of
large $N$ limit should be taken.

It is well known that in the so-called {}'t-Hooft limit \cite{tHooft}
$N\rightarrow \infty$ with $g_{YM}^{2}N$ being fixed, $U(N)$ Yang-Mills
theory simplifies drastically, and some exact analysis, say Borel
summability became possible. Eguchi and Kawai \cite{EK} have argued that
four dimensional large $N$ gauge theory can be replaced by Matrix models
in zero dimensions. In a sense we are studying the reverse process of
the Eguchi-Kawai reduction from a different viewpoint.

In type \twoB matrix models, $g$ appears as the overall factor in the
action and thus there is no nontrivial critical point for $g$. Yet, it
is still controversial \cite{HNT} how large $N$ limit should be taken;
$g$ fixed? $g^{2}N$ fixed? or what else?

In the case of SYM theory embedded in the matrix perturbation theory,
one is faced with an additional complication. Increasing $N$ allows two
different interpretation: (i) larger gauge symmetry (ii) larger cutoff
(i.e. more degrees of freedom).  In the usual QFT case, they are clearly
separated and never mixed.  Here, however, there is a crosstalk between
the two.

In the perturbative computation of $d$-dimensional quantum field theory,
divergences arise from the integrals over loop momenta $\int d^{d}p$. We
can regularize the integrals by introducing a UV cutoff
$\Lambda$. Suppose the amplitude associated with a Feynman diagram
$\Gamma$ diverges as $\Lambda^{D(\Gamma)}$ in the continuum limit
$\Lambda \rightarrow \infty$.  Basically, $D(\Gamma)$ can be determined
from the superficial degree of divergence of $\Gamma$ or its
subgraphs. Renormalizability is the property that all divergence can be
removed if $\Lambda\rightarrow \infty$ limit is taken not keeping $g$
fixed but adjusting $g$ so as to maintain a certain functional relation
${\cal R}(g,\Lambda/\mu)=0$. It is of course a very nontrivial problem
to find the explicit form of renormalization trajectory ${\cal
R}(g,\Lambda/\mu)=0$, but at least perturbatively, it can be determined
by carefully analyzing the structure of divergences in Feynman diagrams.

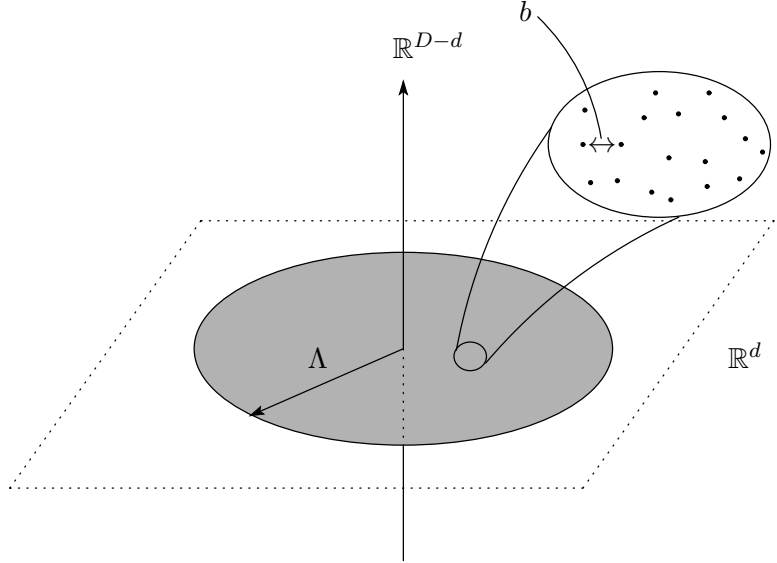
\begin{figure}[tbp]
 \begin{center}
  \unitlength 0.1in
  \begin{picture}(40.00,29.65)(0.00,-31.80)
   %
   \special{pn 8}%
   \special{sh 0.300}%
   \special{ar 2060 2070 1096 506  0.0000000 6.2831853}%
   %
   \special{pn 8}%
   \special{ar 2410 2110 85 75  0.0000000 6.2831853}%
   %
   \special{pn 8}%
   \special{ar 3400 1000 583 383  0.0000000 6.2831853}%
   %
   \special{pn 8}%
   \special{ar 5184 2600 2894 2894  3.3231644 3.7658508}%
   %
   \special{pn 8}%
   \special{ar 4699 4014 2893 2893  3.8436650 4.2859041}%
   %
   \special{pn 8}%
   \special{sh 1}%
   \special{ar 3010 820 10 10 0  6.28318530717959E+0000}%
   \special{sh 1}%
   \special{ar 3040 1200 10 10 0  6.28318530717959E+0000}%
   \special{sh 1}%
   \special{ar 3180 1190 10 10 0  6.28318530717959E+0000}%
   \put(21.9000,-4.8000){\makebox(0,0){$\R^{D-d}$}}%
   %
   \special{pn 8}%
   \special{sh 1}%
   \special{ar 3000 1000 10 10 0  6.28318530717959E+0000}%
   \special{sh 1}%
   \special{ar 3200 1000 10 10 0  6.28318530717959E+0000}%
   \special{sh 1}%
   \special{ar 3360 1250 10 10 0  6.28318530717959E+0000}%
   \special{sh 1}%
   \special{ar 3640 1090 10 10 0  6.28318530717959E+0000}%
   \special{sh 1}%
   \special{ar 3450 1070 10 10 0  6.28318530717959E+0000}%
   \special{sh 1}%
   \special{ar 3460 1290 10 10 0  6.28318530717959E+0000}%
   \special{sh 1}%
   \special{ar 3820 1180 10 10 0  6.28318530717959E+0000}%
   \special{sh 1}%
   \special{ar 3650 1220 10 10 0  6.28318530717959E+0000}%
   \special{sh 1}%
   \special{ar 3660 730 10 10 0  6.28318530717959E+0000}%
   \special{sh 1}%
   \special{ar 3500 840 10 10 0  6.28318530717959E+0000}%
   \special{sh 1}%
   \special{ar 3850 970 10 10 0  6.28318530717959E+0000}%
   \special{sh 1}%
   \special{ar 3740 860 10 10 0  6.28318530717959E+0000}%
   \special{sh 1}%
   \special{ar 3380 730 10 10 0  6.28318530717959E+0000}%
   \special{sh 1}%
   \special{ar 3320 860 10 10 0  6.28318530717959E+0000}%
   \special{sh 1}%
   \special{ar 3940 1040 10 10 0  6.28318530717959E+0000}%
   \put(31.0000,-10.1000){\makebox(0,0){$\leftrightarrow$}}%
   \put(27.0000,-3.0000){\makebox(0,0){$b$}}%
   %
   \special{pn 8}%
   \special{ar 1970 1150 1140 1140  5.4819926 6.1227971}%
   %
   \special{pn 8}%
   \special{pa 2060 2070}%
   \special{pa 2060 670}%
   \special{fp}%
   \special{sh 1}%
   \special{pa 2060 670}%
   \special{pa 2040 737}%
   \special{pa 2060 723}%
   \special{pa 2080 737}%
   \special{pa 2060 670}%
   \special{fp}%
   %
   \special{pn 8}%
   \special{pa 2060 2070}%
   \special{pa 2060 2580}%
   \special{dt 0.045}%
   %
   \special{pn 8}%
   \special{pa 2060 2580}%
   \special{pa 2060 3180}%
   \special{fp}%
   \put(15.6000,-21.7000){\makebox(0,0)[lb]{$\Lambda$}}%
   %
   \special{pn 8}%
   \special{pa 0 2800}%
   \special{pa 0 2800}%
   \special{pa 3000 2800}%
   \special{pa 4000 1400}%
   \special{pa 1000 1400}%
   \special{pa 1000 1400}%
   \special{pa 0 2800}%
   \special{dt 0.045}%
   %
   \special{pn 8}%
   \special{pa 2060 2070}%
   \special{pa 1260 2420}%
   \special{fp}%
   \special{sh 1}%
   \special{pa 1260 2420}%
   \special{pa 1329 2412}%
   \special{pa 1309 2399}%
   \special{pa 1313 2375}%
   \special{pa 1260 2420}%
   \special{fp}%
   \put(37.6000,-21.7000){\makebox(0,0)[lb]{$\R^d$}}%
  \end{picture}%
 \end{center}
 \caption{D-instanton configuration corresponding to $d$-dimensional
 SYM. $N$ D-instantons are distributed in a $d$-dimensional ball with
 radius $\Lambda$ within $\R^{d}$} \label{fig:SYMfiniteN}
\end{figure}

On the other hand, in the matrix perturbation theory we are working
with, only source of divergence is sending $N$ to infinity.\footnote{Of
course, there are combinatorial divergences due to the infinitely many
Feynman graphs. But this is common to Matrix and to both theories and is
not discussed in this paper.}  In the spirit of correspondence between
Matrix model and QFT, continuum limit $\Lambda\rightarrow \infty$ should
be related to the large $N$ limit.  In other words, adding more and more
D-instantons on the outskirts of the D-instanton cluster should be
equivalent to increasing $\Lambda$.  In this picture, the UV cutoff
$\Lambda$ is nothing but the distance to the {\it farthest} D-instanton
(see Fig.~\ref{fig:finiteN}.), which is natural from UV/IR
correspondence \cite{SW} or spacetime uncertainty \cite{Yoneya}.

The analysis of large $N$ behavior can be complicated because there is
no unique way to add extra D-instantons; the relation between the two
limits $N\rightarrow \infty$ and $\Lambda\rightarrow \infty$ is highly
dependent on the strategy of putting new D-instantons.

In this section, we will take a phenomenological approach to clarify
the relation between large $N$ limit and continuum limit
$\Lambda\rightarrow \infty$. $N$ D-instantons are assumed to be
concentrated along $d$-dimensional hyperplane with uniform density
$\rho$ as in section \ref{sec:ddimSYM}. But here, since $N$ is finite,
the radius $\Lambda$ of D-instanton cluster is also finite.

  \subsection{Scaling laws}
\label{sec:Scaling}

So far we have encountered various length scales. The core size
$g^{1/2}$, the D-instanton spacing $a$, the scale of external momenta
$\mu$. For finite $N$, the distance $\Lambda$ to the farthest
D-instanton will also play an important role.  We will study the scaling
laws for these length scales. For the readers' convenience they are
listed in Table 1.  Although the word ``length'' will be frequently
used, they represent momentum scales in the QFT picture as we argued in
the previous section.

\begin{table}[tbp]
 \centering
 \renewcommand{\arraystretch}{1.2}
 \begin{tabular}{@{\vrule width 1pt\ \ }l|l|l|l@{\ \vrule width 1pt}}
  \bhline
  $g^{1/2}$ & $\sim N^{-\omc/4}$ 
  & core size & the minimum distance between D-instantons\\
  \hline
  $a$  &  $\sim N^{-\om/d}$ 
  & D-instanton spacing & 
  average distance to the nearest D-instantons
  \\
  \hline
  $\mu$ & $\sim N^{0}$ 
  &  renormalization scale
  & typical momentum scale of the external lines
  \\
  \hline
  $\Lambda$ & $\sim N^{(1-\om)/d}$ 
  & cutoff & distance to the farthest D-brane\\
  \bhline
 \end{tabular}
 \caption{Various scales in D-instanton distribution}
 \label{tab:LinearScales}
\end{table}

\begin{table}[tbp]
 \centering
 \renewcommand{\arraystretch}{1.2}
 \begin{tabular}{@{\vrule width 1pt\ \ }l|l|l|l@{\ \vrule width 1pt}}
  \bhline
  $b$  & $\sim N^{-\frac{\om}{d}}$ & 
  $\sim \Lambda^{\frac{-\om}{1-\om}}$ 
  & cluster spacing \\
  \hline
  $\rho$  & $\sim N^{\om}$ & 
  $\sim \Lambda^{\frac{d\om}{1-\om}}$ 
  & D-instanton density  \\
  \hline
  $\rho'$  & $\sim N^{\om}$ & 
  $\sim \Lambda^{\frac{d\om}{1-\om}}$ 
  & cluster density  \\
  \hline
  $g_{d}^2$ & $\sim N^{\om-\omc}$ & 
  $\sim \Lambda^{\frac{d(\om-\omc)}{1-\om}}$ 
  & YM coupling in $d$-dimensional QFT
  \\
  \bhline
 \end{tabular}
 \caption{Other quantities with nontrivial scaling laws}
 \label{tab:OtherScales}
\end{table}

We are interested in how they should be varied as $N$ tends to infinity. 
In this paper, we assume a simple power law scaling and try to draw some
bounds on the exponents from physically reasonable assumptions.

Since the quantities in Table \ref{tab:LinearScales} are all
dimensionful while $N$ is dimensionless, one must decide which is kept
fixed in the large $N$ limit. For this purpose we choose $\mu$, the
momentum scale carried by the external lines in the QFT picture.  In
other words, all ``lengths'' discussed in this section are measured in
the unit of $\mu$.  For example, we set $\mu \sim x_{ij}$ when we
compute $\langle X_{ij}^{\mu} X_{ji}^{\nu} \rangle$.  The other three
quantities $g^{1/2}, a$ and $\Lambda$ are assumed to scale with some
power of $N$ which is specified by three independent exponents $d$,
$\om$ and $\omc$ as in Table \ref{tab:LinearScales}.

In addition to these basic ``length'' scales, there are some other
quantities of interest, with nontrivial $N$ dependence:
\begin{itemize}
 \item {D-instanton density $\rho$}
       
       For finite $N$, D-instantons are assumed to be distributed
       uniformly with density $\rho$ within a $d$-dimensional ball of
       radius $\Lambda$. Thus we have
       \begin{equation}
        \rho \Lambda^{d} \sim N 
         \label{rhoLambdaN}
       \end{equation}
       Since $\Lambda \sim N^{(1-\om)/d}$, (\ref{rhoLambdaN})
       fixes the scaling of the $\rho$ as
       \begin{equation}
        \rho \sim N^{\om}.
         \label{densn} 
       \end{equation}
       
 \item {cluster spacing $b$}
       
       The D-instanton spacing $a$ and cluster size $b$ are related via $n
       a^{d}=b^{d}$. Since we fix the rank $n$ of the gauge group, $a$
       and $b$ will have the same scaling behavior, $b\sim a\sim
       N^{-\om/d}$. 
       
 \item {cluster density $\rho'$}
       
       By the same token, the cluster density $\rho'$, related to
       D-instanton density $\rho$ via $n\rho'=\rho$ will have the same
       scaling as $\rho$.
       \begin{equation}
        \rho ' \sim N^{\om}
       \end{equation}

 \item {YM coupling in $d$-dimensions $g_{d}$}
       
       YM coupling  $g_{d}$ in $d$-dimensions is related to 
       $g$ and $\rho'$ via (\ref{gddef}). Thus it will have a
       nontrivial
       $N$ dependence:
       \begin{equation}
        g_{d}^{2} \sim \rho' g^{2} \sim N^{\om-\omc}
       \end{equation}
\end{itemize}
We summarize the result in Table \ref{tab:OtherScales}.

  \subsection{Physical bounds on scaling exponents}
\label{subsec:bounds}

We have seen that a $d$-dimensional SYM theory emerges from the
off-diagonal dynamics of the large $N$ Matrix model. In order to prove
the claim, we need at least to show such a large $N$ limit is indeed
possible --- precisely specifying how to arrange the background
D-instanton configuration as $N$ tends to infinity. It may be difficult
to do this rigorously.  We will content ourselves with obtaining some
inequalities among the exponents $\om$, $\omc$, $d$ introduced in
section \ref{sec:Scaling}, so that there occurs no apparent inconsistency
in QFT side. This would help us applying Matrix theory to more realistic
situations in the future.

We will consider several physically reasonable assumptions, but we do
not intend to claim that following conditions are all necessary or
sufficient.

{\sf(i)}\quad Interpretation as $d$-dimensional QFT required replacing
the discrete sum by $d$-dimensional integral. This coarse graining can
be justified only when $a \sim b \ll \mu$. This is true in the large $N$
limit if
\begin{equation}
 \om \geq 0.
  \label{PositiveOmega}
\end{equation}

{\sf(ii)} \quad From the $d$-dimensional point of view, cutoff scale
$\Lambda$ must tend to infinity. Thus we have
\begin{equation}
 \om \leq 1.
\end{equation}

{\sf(iii)} \quad As we saw in section \ref{sec:Perturbation}, the
perturbation theory is essentially the expansion in $g/a^{2}$. Thus it
is valid if $a \gg g^{1/2}$ is satisfied. This remains to be true in
large $N$ limit if
\begin{equation}
 d \, \omc \ge 4 \, \om.
  \label{PerturbationBound}
\end{equation}

{\sf(iv)} \quad Actually, the same bound can be obtained from a
different viewpoint. From exact results for matrix integrals
\cite{Piljin,Savdeep,Suyama}, it is reasonable to assume there is a
pairwise repulsive potential among D-instantons due to entropy factor.
This could be effectively treated \cite{AIKKT} as each D-instanton has a
core size of order $g^{1/2}$. This implies $a \gtrsim g^{1/2}$. Sending
$N$ to infinity, we obtain the inequality (\ref{PerturbationBound}).

{\sf(v)} \quad To construct $U(n)$ gauge theory, $n$ D-instantons are
put on the same point (see section \ref{sec:ddimSYM}). But this
assumption might be too strong; it is possible that the $n$ D-instantons
can disperse in the cluster of size $b$ but are still grouped via
(slightly broken) $U(n)$ gauge action from QFT point of
view.\footnote{This claim is not so strange as it sounds. In Nature,
non-Abelian symmetry is exact in UV regime but hidden in IR regime
through confinement or Higgs mechanism. In our context, the D-instantons
within a cluster look almost coincident in much larger scale $\mu$.} Of
course, the dispersion size $b$ must be sufficiently smaller than the
renormalization scale,
\begin{equation}
 b \ll \mu .\label{bllm}
\end{equation}
This corresponds to the minimum momentum resolution seen by $U(n)$
Yang-Mills theory. The condition (\ref{bllm}) leads to the bound,
(\ref{PositiveOmega}).

{\sf(vi)} \quad As far as a tree level amplitude or the form of
Yang-Mills action is concerned, the argument given in {\sf(v)} is
sufficient. But if quantum effects are taken into account, it is another
story.  Just like the anomaly from one loop, large $N$ divergence from
loop integrals may overwhelm the $b/\mu \sim N^{-\om}$ suppression
discussed in {\sf(v)} and may yield non-negligible effects.

Let us estimate the effect of dispersing D-instantons using the one-loop
two point function as an example.  As a function of D-instanton
configuration $\{x_{k}\}$, the most divergent contribution is roughly
given by
\begin{equation}
 \begin{split}
  {\cal A}^{\text{1-loop}}[\{x_{k}\}]
  &
  \sim \frac{g^{2}}{\mu^{2}}\sum_{k}\frac{1}{x_{ik}^{2}}
  \sim \frac{g^{2}\rho}{\mu^{2}}
  \int^{\Lambda}\frac{d^{d}p}{p^{2}+\mu^{2}}
  \\
  &
  \sim \frac{g^{2}\rho}{\mu^{2}}\Lambda^{d-2}
 \end{split}
\end{equation}
If the D-instanton positions $\{x_{k}\}$ have a dispersion of order $b$,
$ {\cal A}^{\text{1-loop}}[\{x_{k}\}]$ will change as
\begin{equation}
 \begin{split}
  \delta {\cal A}^{\text{1-loop}} & = 
  {\cal A}^{\text{1-loop}}[\{x_{k}+\delta x_{k}\}]
  -
  {\cal A}^{\text{1-loop}}[\{x_{k}\}]
  \\
  &
  \sim \frac{g^{2}}{\mu^{2}}\sum_{k}
  \left\{
  \frac{1}{(x_{ik}-\delta x_{k})^{2}}
  - \frac{1}{x_{ik}^{2}}
  \right\}
  \\
  & 
  \lesssim \frac{g^{2}\rho}{\mu^{2}}
  \int^{\Lambda} d^{d}p \;
  b \frac{\partial}{\partial p}
  \left(\frac{1}{p^{2}+\mu^{2}}\right)
  \sim \frac{g^{2}\rho}{\mu^{2}} b \Lambda^{d-3}.
 \end{split}
\end{equation}
The last expression scales as $N$ to the
$(\om-\omc-\frac{\om}{d}+(d-3)\frac{1-\om}{d})$-th power. 
Thus, it is negligible only if
\begin{equation}
 d (1-\omc) < 3-2\om
\end{equation}
This is the condition when the operators associated with the slight
shift of D-instantons are irrelevant in the sense of large $N$
renormalization group.  \label{item:OneLoop}

{\sf(vii)} \quad Let us study the problem of renormalizability of
$d$-dimensional theory i.e. whether or not as $N\rightarrow \infty$ only
a finite number of amplitudes superficially diverge.

From the Wilsonian point of view, renormalizability is not a necessary
condition for QFT, but a consequence of the renormalization
procedure. But it is of some interest in presenting the analysis since
in the context of Matrix model, situation is rather
complicated.\footnote{For example, it is not clear ``What is IR
limit?''}

For a given Feynman graph, the superficial degree of divergence is
usually determined from the number of loops and propagators. The
coupling constants just count the number of vertices and stay fixed when
the cutoff is sent to infinity. As we all know, $d=4$ is the critical
dimension for gauge theories.

But here, we are talking about the divergence when $N$ tends to
infinity. Recall not only $\Lambda$ but also $g_{d}$ changes as a
function of $N$, i.e.  $g_{d}^{2}\sim N^{\om-\omc}$. Thus usual
definition of the superficial degree of divergence does not work.

In a sense, we are studying a generalized large $N$ limit in which
$g_{d}^{2} N^{\omc-\om}$ is kept fixed. Thus the standard QFT results
should follow if we restrict to $\om=\omc$, whereas 't Hooft limit would
correspond to another special case, $\om+1=\omc$.

What is the new rule for the superficial degree of divergence?  Note
that $g^2$ always come in pair with a sum over D-instantons. From the
substitution
\begin{equation}
 g^{2}\sum_{k} \cdots 
  \Longrightarrow g^{2}  \rho' \int d^{d}p \tr (\cdots)
  \Longrightarrow g_{d}^{2} \int d^{d}p \tr (\cdots),
\end{equation}
extra $d$-dimensional loop integral is always associated with the factor
$g_{d}^2$. Each loop contributes
$\Lambda^{d}$ while the coupling gives $g_{d}^{2}\sim
N^{\om-\omc}\sim
\Lambda^{d(\om-\omc)/(1-\om)}$. It is easy to
convince oneself that the net effect is to replace the spacetime
dimension $d$ by an effective dimension
\begin{equation}
 d_{\textit{eff}} \equiv d + \frac{d(\om-\omc)}{1-\om}
  = d  \frac{1-\omc}{1-\om}.
\end{equation}
Thus we have a new criteria about large $N$ renormalizability as
follows
\begin{equation}
 d_{\textit{eff}}\leq 4 \Longleftrightarrow 
  d (1-\omc) \le 4 (1-\om).
\end{equation}
As promised, $\om=\omc$ recovers the standard result.  Note that if
$\omc=1$, $d_{\textit{eff}}=0$ for any $d, \om$.  This corresponds to
the well known fact that planar limit of the $0+0$-dimensional Matrix
model absolutely converge.

{\sf(viii)} \quad As for the error in replacing sums by integrals,
analysis in {\sf(vii)} can be generalized to an arbitrary Feynman graph
$\Gamma$.  Suppose we know the amplitude diverges as \[ {\cal A}(\Gamma)
\sim N^{\gamma(\Gamma)}, \] including $N$ dependence of $g_{d}^{2}$.
Then, the approximation can be justified if 
\begin{equation*}
 \delta {\cal A} \sim
  \frac{b}{\Lambda}N^{\gamma(\Gamma)}\sim N^{\gamma(\Gamma)-\frac{1}{d}}
  \ll 1.
\end{equation*}
Thus the error is negligible for graphs with sufficiently low
degree of divergence:
\begin{equation}
 \gamma(\Gamma) < \frac{1}{d}.
  \label{gang}
\end{equation}
It may be useful to introduce effective superficial degree of
divergence, ${\cal D}_{\textit{eff}}(\Gamma)$ defined
through\footnote{${\cal D}_{\textit{eff}}(\Gamma)$ coincides with usual
superficial degree of divergence ${\cal D}(\Gamma)$ if $g_{d}$ is
independent of $N$.}  \[ \Lambda ^{{\cal D}_{\textit{eff}}(\Gamma)} \sim
N^{\gamma(\Gamma)}, \] or equivalently \[ {\cal
D}_{\textit{eff}}(\Gamma) = \frac{d}{1-\om}\gamma(\Gamma).  \] Then,
(\ref{gang}) can be expressed as
\begin{equation}
 {\cal D}_{\textit{eff}}(\Gamma) < \frac{1}{1-\om}
  \Longleftrightarrow 
  \om > 1-\frac{1}{{\cal D}_{\textit{eff}}(\Gamma)} \label{eff}
\end{equation}
Suppose $d$-dimensional theory is renormalizable in the sense that
${\cal D}_{\textit{eff}}(\Gamma)$ has a $\Gamma$-independent upper bound
${\cal D}_{\textit{max}}$. Then by choosing
\begin{equation*}
 \om >
 1-\frac{1}{{\cal D}_{\textit{max}}}, 
\end{equation*}
the error can be neglected for all Feynman integrals.  In particular,
$d=4$ SYM case (${\cal D}_{\textit{max}} = 2 $) leaves as a finite
window $1>\om > \frac{1}{2}$.

If $d$-dimensional theory is non-renormalizable, ${\cal
D}_{\textit{eff}}$ grows with the number $L$ of loops.  Thus from
(\ref{eff}), for a given $\om$, there is a maximal number of loops
$L_{\textit{max}}$ beyond which the approximation fails.  But
$L_{\textit{max}}$ tends to infinity if we approach {}'t Hooft limit,
$\om \to 1$.

\subsection{Possible interpretation of exponents}

So far we have chosen a particular D-instanton configuration depicted in
Fig \ref{fig:SYMfiniteN} the dimension $d$ in which QFT lives is
determined by the number of flat directions.  But this is clearly a very
special configuration from $D$-dimensional viewpoint.  Can we relax the
assumption?

In Wilson's approach to renormalization group, one can study the origin
of ultraviolet divergences by isolating the dependence of the functional
integral on the short distance degrees of freedom of the field.  In
Matrix approach, short distance degrees of freedom correspond to the
long distance D-instantons. The number $\delta N$ of D-instantons
contained in the momentum shell $\Lambda < |p|< \Lambda+ \delta \Lambda
$ is given by
\begin{equation*}
 \delta N \propto \Lambda ^{d-1}\delta \Lambda \qquad
 \text{as} \quad \Lambda \rightarrow \infty.
\end{equation*}
In fact, as far as the loop divergence is concerned,
dimension $d$ will appear only through this relation.

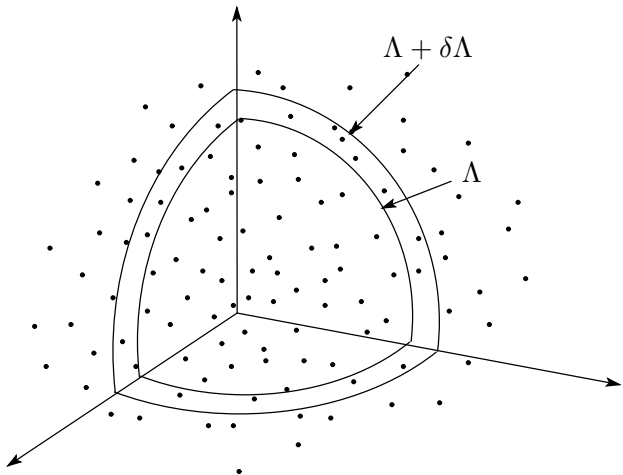
\begin{figure}[tbp]
 \begin{center}
  \unitlength 0.1in
  \begin{picture}(32.03,24.30)(2.10,-28.20)
   %
   \special{pn 8}%
   \special{ar 1453 1633 1133 787  0.7065328 2.0752866}%
   %
   \special{pn 8}%
   \special{ar 1384 2006 946 1036  4.7620310 6.2831853}%
   \special{ar 1384 2006 946 1036  0.0000000 0.1293807}%
   %
   \special{pn 8}%
   \special{ar 1990 2140 1102 1366  3.0052792 4.1641425}%
   %
   \special{pn 8}%
   \special{ar 1423 1596 1355 924  0.7070358 2.0746233}%
   %
   \special{pn 8}%
   \special{ar 1338 2035 1132 1217  4.7626998 6.2831853}%
   \special{ar 1338 2035 1132 1217  0.0000000 0.1277815}%
   %
   \special{pn 8}%
   \special{ar 2077 2189 1317 1604  3.0051303 4.1642380}%
   %
   \special{pn 8}%
   \special{pa 1410 1990}%
   \special{pa 1409 390}%
   \special{fp}%
   \special{sh 1}%
   \special{pa 1409 390}%
   \special{pa 1389 457}%
   \special{pa 1409 443}%
   \special{pa 1429 457}%
   \special{pa 1409 390}%
   \special{fp}%
   %
   \special{pn 8}%
   \special{pa 1410 1990}%
   \special{pa 210 2791}%
   \special{fp}%
   \special{sh 1}%
   \special{pa 210 2791}%
   \special{pa 277 2771}%
   \special{pa 254 2761}%
   \special{pa 254 2737}%
   \special{pa 210 2791}%
   \special{fp}%
   %
   \special{pn 8}%
   \special{pa 1410 1990}%
   \special{pa 3413 2366}%
   \special{fp}%
   \special{sh 1}%
   \special{pa 3413 2366}%
   \special{pa 3351 2334}%
   \special{pa 3361 2356}%
   \special{pa 3344 2373}%
   \special{pa 3413 2366}%
   \special{fp}%
   \put(26.4000,-12.5000){\makebox(0,0){$\Lambda$}}%
   \put(24.1000,-6.0000){\makebox(0,0){$\Lambda+\delta \Lambda$}}%
   %
   \special{pn 8}%
   \special{pa 2360 690}%
   \special{pa 2000 1050}%
   \special{fp}%
   \special{sh 1}%
   \special{pa 2000 1050}%
   \special{pa 2061 1017}%
   \special{pa 2038 1012}%
   \special{pa 2033 989}%
   \special{pa 2000 1050}%
   \special{fp}%
   %
   \special{pn 8}%
   \special{pa 2530 1300}%
   \special{pa 2170 1440}%
   \special{fp}%
   \special{sh 1}%
   \special{pa 2170 1440}%
   \special{pa 2239 1434}%
   \special{pa 2220 1421}%
   \special{pa 2225 1397}%
   \special{pa 2170 1440}%
   \special{fp}%
   %
   \special{pn 8}%
   \special{sh 1}%
   \special{ar 2300 740 10 10 0  6.28318530717959E+0000}%
   \special{sh 1}%
   \special{ar 2280 980 10 10 0  6.28318530717959E+0000}%
   \special{sh 1}%
   \special{ar 2010 1040 10 10 0  6.28318530717959E+0000}%
   \special{sh 1}%
   \special{ar 2000 810 10 10 0  6.28318530717959E+0000}%
   \special{sh 1}%
   \special{ar 1650 810 10 10 0  6.28318530717959E+0000}%
   \special{sh 1}%
   \special{ar 1690 960 10 10 0  6.28318530717959E+0000}%
   \special{sh 1}%
   \special{ar 1920 1020 10 10 0  6.28318530717959E+0000}%
   \special{sh 1}%
   \special{ar 1710 1160 10 10 0  6.28318530717959E+0000}%
   \special{sh 1}%
   \special{ar 1720 1870 10 10 0  6.28318530717959E+0000}%
   \special{sh 1}%
   \special{ar 1720 2080 10 10 0  6.28318530717959E+0000}%
   \special{sh 1}%
   \special{ar 350 2060 10 10 0  6.28318530717959E+0000}%
   \special{sh 1}%
   \special{ar 410 2270 10 10 0  6.28318530717959E+0000}%
   \special{sh 1}%
   \special{ar 760 1910 10 10 0  6.28318530717959E+0000}%
   \special{sh 1}%
   \special{ar 430 1650 10 10 0  6.28318530717959E+0000}%
   \special{sh 1}%
   \special{ar 680 1310 10 10 0  6.28318530717959E+0000}%
   \special{sh 1}%
   \special{ar 1070 1010 10 10 0  6.28318530717959E+0000}%
   \special{sh 1}%
   \special{ar 930 910 10 10 0  6.28318530717959E+0000}%
   \special{sh 1}%
   \special{ar 1430 980 10 10 0  6.28318530717959E+0000}%
   \special{sh 1}%
   \special{ar 1380 1360 10 10 0  6.28318530717959E+0000}%
   \special{sh 1}%
   \special{ar 1730 2680 10 10 0  6.28318530717959E+0000}%
   \special{sh 1}%
   \special{ar 1900 2540 10 10 0  6.28318530717959E+0000}%
   \special{sh 1}%
   \special{ar 1420 2820 10 10 0  6.28318530717959E+0000}%
   \special{sh 1}%
   \special{ar 1390 2580 10 10 0  6.28318530717959E+0000}%
   \special{sh 1}%
   \special{ar 1260 2580 10 10 0  6.28318530717959E+0000}%
   \special{sh 1}%
   \special{ar 1740 2530 10 10 0  6.28318530717959E+0000}%
   \special{sh 1}%
   \special{ar 2470 2460 10 10 0  6.28318530717959E+0000}%
   \special{sh 1}%
   \special{ar 2280 2270 10 10 0  6.28318530717959E+0000}%
   \special{sh 1}%
   \special{ar 2020 2350 10 10 0  6.28318530717959E+0000}%
   \special{sh 1}%
   \special{ar 2220 2470 10 10 0  6.28318530717959E+0000}%
   \special{sh 1}%
   \special{ar 2620 2250 10 10 0  6.28318530717959E+0000}%
   \special{sh 1}%
   \special{ar 2680 1880 10 10 0  6.28318530717959E+0000}%
   \special{sh 1}%
   \special{ar 2520 1980 10 10 0  6.28318530717959E+0000}%
   \special{sh 1}%
   \special{ar 2730 2050 10 10 0  6.28318530717959E+0000}%
   \special{sh 1}%
   \special{ar 2920 1810 10 10 0  6.28318530717959E+0000}%
   \special{sh 1}%
   \special{ar 2280 1140 10 10 0  6.28318530717959E+0000}%
   \special{sh 1}%
   \special{ar 2620 1100 10 10 0  6.28318530717959E+0000}%
   \special{sh 1}%
   \special{ar 2570 1380 10 10 0  6.28318530717959E+0000}%
   \special{sh 1}%
   \special{ar 2440 1240 10 10 0  6.28318530717959E+0000}%
   \special{sh 1}%
   \special{ar 2880 1410 10 10 0  6.28318530717959E+0000}%
   \special{sh 1}%
   \special{ar 2830 1550 10 10 0  6.28318530717959E+0000}%
   \special{sh 1}%
   \special{ar 2230 1790 10 10 0  6.28318530717959E+0000}%
   %
   \special{pn 8}%
   \special{sh 1}%
   \special{ar 1380 2270 10 10 0  6.28318530717959E+0000}%
   \special{sh 1}%
   \special{ar 1550 2180 10 10 0  6.28318530717959E+0000}%
   \special{sh 1}%
   \special{ar 1910 2260 10 10 0  6.28318530717959E+0000}%
   \special{sh 1}%
   \special{ar 1500 2440 10 10 0  6.28318530717959E+0000}%
   \special{sh 1}%
   \special{ar 1560 2240 10 10 0  6.28318530717959E+0000}%
   \special{sh 1}%
   \special{ar 1230 1770 10 10 0  6.28318530717959E+0000}%
   \special{sh 1}%
   \special{ar 1440 1560 10 10 0  6.28318530717959E+0000}%
   \special{sh 1}%
   \special{ar 1630 1490 10 10 0  6.28318530717959E+0000}%
   \special{sh 1}%
   \special{ar 1580 1700 10 10 0  6.28318530717959E+0000}%
   \special{sh 1}%
   \special{ar 1470 1910 10 10 0  6.28318530717959E+0000}%
   \special{sh 1}%
   \special{ar 2060 1930 10 10 0  6.28318530717959E+0000}%
   \special{sh 1}%
   \special{ar 2090 2090 10 10 0  6.28318530717959E+0000}%
   \special{sh 1}%
   \special{ar 2190 2030 10 10 0  6.28318530717959E+0000}%
   \special{sh 1}%
   \special{ar 2140 1590 10 10 0  6.28318530717959E+0000}%
   \special{sh 1}%
   \special{ar 1930 1640 10 10 0  6.28318530717959E+0000}%
   \special{sh 1}%
   \special{ar 1950 1760 10 10 0  6.28318530717959E+0000}%
   \special{sh 1}%
   \special{ar 1870 1820 10 10 0  6.28318530717959E+0000}%
   \special{sh 1}%
   \special{ar 1850 1410 10 10 0  6.28318530717959E+0000}%
   \special{sh 1}%
   \special{ar 1960 1370 10 10 0  6.28318530717959E+0000}%
   \special{sh 1}%
   \special{ar 1520 1120 10 10 0  6.28318530717959E+0000}%
   \special{sh 1}%
   \special{ar 1530 1300 10 10 0  6.28318530717959E+0000}%
   \special{sh 1}%
   \special{ar 1730 1290 10 10 0  6.28318530717959E+0000}%
   \special{sh 1}%
   \special{ar 1380 1280 10 10 0  6.28318530717959E+0000}%
   \special{sh 1}%
   \special{ar 1250 1450 10 10 0  6.28318530717959E+0000}%
   \special{sh 1}%
   \special{ar 1090 1710 10 10 0  6.28318530717959E+0000}%
   \special{sh 1}%
   \special{ar 1060 2050 10 10 0  6.28318530717959E+0000}%
   \special{sh 1}%
   \special{ar 1390 1950 10 10 0  6.28318530717959E+0000}%
   \special{sh 1}%
   \special{ar 1290 2010 10 10 0  6.28318530717959E+0000}%
   \special{sh 1}%
   \special{ar 1140 2230 10 10 0  6.28318530717959E+0000}%
   \special{sh 1}%
   \special{ar 1110 2420 10 10 0  6.28318530717959E+0000}%
   \special{sh 1}%
   \special{ar 900 2540 10 10 0  6.28318530717959E+0000}%
   \special{sh 1}%
   \special{ar 880 2270 10 10 0  6.28318530717959E+0000}%
   \special{sh 1}%
   \special{ar 970 1780 10 10 0  6.28318530717959E+0000}%
   \special{sh 1}%
   \special{ar 1150 1910 10 10 0  6.28318530717959E+0000}%
   \special{sh 1}%
   \special{ar 1170 1500 10 10 0  6.28318530717959E+0000}%
   \special{sh 1}%
   \special{ar 1320 1600 10 10 0  6.28318530717959E+0000}%
   \special{sh 1}%
   \special{ar 1120 1230 10 10 0  6.28318530717959E+0000}%
   \special{sh 1}%
   \special{ar 1010 1410 10 10 0  6.28318530717959E+0000}%
   \special{sh 1}%
   \special{ar 1270 1170 10 10 0  6.28318530717959E+0000}%
   \special{sh 1}%
   \special{ar 830 1620 10 10 0  6.28318530717959E+0000}%
   \special{sh 1}%
   \special{ar 690 1570 10 10 0  6.28318530717959E+0000}%
   \special{sh 1}%
   \special{ar 1000 1250 10 10 0  6.28318530717959E+0000}%
   \special{sh 1}%
   \special{ar 1220 800 10 10 0  6.28318530717959E+0000}%
   \special{sh 1}%
   \special{ar 540 2050 10 10 0  6.28318530717959E+0000}%
   \special{sh 1}%
   \special{ar 660 2200 10 10 0  6.28318530717959E+0000}%
   \special{sh 1}%
   \special{ar 1760 2240 10 10 0  6.28318530717959E+0000}%
   \special{sh 1}%
   \special{ar 2360 1600 10 10 0  6.28318530717959E+0000}%
   \special{sh 1}%
   \special{ar 2360 1770 10 10 0  6.28318530717959E+0000}%
   \special{sh 1}%
   \special{ar 2180 1350 10 10 0  6.28318530717959E+0000}%
   \special{sh 1}%
   \special{ar 1960 1080 10 10 0  6.28318530717959E+0000}%
   \special{sh 1}%
   \special{ar 2030 1180 10 10 0  6.28318530717959E+0000}%
   \special{sh 1}%
   \special{ar 2480 1570 10 10 0  6.28318530717959E+0000}%
   \special{sh 1}%
   \special{ar 2500 1700 10 10 0  6.28318530717959E+0000}%
   \special{sh 1}%
   \special{ar 750 2520 10 10 0  6.28318530717959E+0000}%
   \special{sh 1}%
   \special{ar 620 2380 10 10 0  6.28318530717959E+0000}%
   \special{sh 1}%
   \special{ar 850 1400 10 10 0  6.28318530717959E+0000}%
   \special{sh 1}%
   \special{ar 1310 1010 10 10 0  6.28318530717959E+0000}%
   \special{sh 1}%
   \special{ar 1520 730 10 10 0  6.28318530717959E+0000}%
   \special{sh 1}%
   \special{ar 1800 1650 10 10 0  6.28318530717959E+0000}%
   %
   \special{pn 8}%
   \special{sh 1}%
   \special{ar 1490 1770 10 10 0  6.28318530717959E+0000}%
   \special{sh 1}%
   \special{ar 1600 1930 10 10 0  6.28318530717959E+0000}%
   %
   \special{pn 8}%
   \special{sh 1}%
   \special{ar 1450 2090 10 10 0  6.28318530717959E+0000}%
   \special{sh 1}%
   \special{ar 1330 1820 10 10 0  6.28318530717959E+0000}%
   \special{sh 1}%
   \special{ar 1330 2150 10 10 0  6.28318530717959E+0000}%
   \special{sh 1}%
   \special{ar 1870 1950 10 10 0  6.28318530717959E+0000}%
   \special{sh 1}%
   \special{ar 1620 1780 10 10 0  6.28318530717959E+0000}%
   %
   \special{pn 8}%
   \special{sh 1}%
   \special{ar 940 1980 10 10 0  6.28318530717959E+0000}%
   \special{sh 1}%
   \special{ar 940 1580 10 10 0  6.28318530717959E+0000}%
   \special{sh 1}%
   \special{ar 810 2140 10 10 0  6.28318530717959E+0000}%
   \special{sh 1}%
   \special{ar 600 1790 10 10 0  6.28318530717959E+0000}%
   \special{sh 1}%
   \special{ar 870 1190 10 10 0  6.28318530717959E+0000}%
   \special{sh 1}%
   \special{ar 1670 2390 10 10 0  6.28318530717959E+0000}%
   \special{sh 1}%
   \special{ar 1240 2340 10 10 0  6.28318530717959E+0000}%
  \end{picture}%
 \end{center}
 \caption{General D-instanton configuration. The number
 $\delta N$ of D-instantons contained in the shell
 $\Lambda<|p|<\Lambda+\delta\Lambda$ is proportional to
 $\Lambda^{d-1}$.}
 \label{fig:Globular} 
\end{figure}

Consider for example a more generic D-instanton configuration, Fig
\ref{fig:Globular}. Let $N(\Lambda)$ be the number of D-instantons
within a $D$-dimensional ball of radius $\Lambda$.  Then the space-time
dimension can be ``defined'' as the rate of growth of $N$:
\begin{equation}
 d = \frac{\partial \log N(\Lambda)}{\partial \log \Lambda}. 
  \label{loglog}
\end{equation}
For a uniform configuration like Fig \ref{fig:SYMfiniteN}, definition
(\ref{loglog}) gives the number of flat directions.  Note that the new
definition (\ref{loglog}) make sense for non-integer dimension $d$,
which may be useful to visualize the meaning of dimensional
regularization.

In order to enumerate the physical degrees of freedom in a field theory,
one needs to put the system into a finite box of volume $V$. The number
of states is given by the available phase space volume.
\begin{equation}
 N = \frac{V}{(2 \pi \hbar)^{d}} \int^{\Lambda} d^{d}p \; .
  \label{NVint}
\end{equation}
In order to realize a continuum field theory in infinite spacetime,
one needs to take two limits:
\begin{equation*}
 \begin{split}
  V\rightarrow \infty & \qquad  (\text{Large volume limit}), \\
  \Lambda\rightarrow \infty & \qquad (\text{Continuum limit}).  
 \end{split}
\end{equation*}
Singularities associated with the former and latter are usually called
IR and UV divergences, respectively.  In standard textbooks on QFT,
$V\rightarrow \infty$ limit is taken first so that Feynman rules
simplify in the momentum space. Subsequently, $\Lambda\rightarrow
\infty$ limit is carefully investigated.  This asymmetry between the two
limits is due to the well known fact that the translational invariance
in momentum space is actually broken by the hierarchical structure.

In the Matrix model, however, all limiting processes are ``unified''
into a single large $N$ limit. Comparing (\ref{SubsSummary}) and
(\ref{NVint}), we can say that we have investigated in section
\ref{subsec:bounds} all possible limits
\begin{equation}
 V \sim \rho' \sim N^{\om}, \qquad  \Lambda \sim N^{(1-\om)/d}
\end{equation}
to get a continuum field theory.

 \section{Discussions}
\label{sec:Discussion}

In this paper, we have studied the dynamics of large $N$ Matrix models
through the quantum fluctuations in a fixed D-instanton background.  In
particular, we have explicitly shown the correspondence of perturbation
theories between the usual QFT and Matrix perturbation theory. The
correspondence is exact if relative D-instanton positions are
interpreted as momenta in QFT picture.

One might think that this is a kind of triviality.  Indeed, Matrix model
action is originated from the Yang-Mills action by dimensional
reduction.  It is no wonder Yang-Mills theory can be recovered from the
Matrix model. However, since dimensional reduction is simply throwing
away spacetime coordinate dependence, the reverse procedure would be
just re-introducing $x$ dependence to the matrix fields. But contrary to
this naive expectation, the momentum space picture emerges first and
coordinate picture is recovered only after Fourier transformation.

The correspondence exploited in this paper can be regarded as a ``dual''
version of Eguchi-Kawai reduction \cite{EK}.  The original suggestion by
Eguchi and Kawai is valid only at strong coupling \cite{BHN}, whereas we
have shown the equivalence in a weak coupling regime.

Just like lattice gauge theories, Matrix model provides us with a
natural gauge invariant regularization. But ``Matrix regularization''
has two important features.  First, the quantum fields are discretized
in the momentum space rather than ordinary space, and the hierarchical
structure inherent to QFT can be understood in a geometrical fashion.
Second, matrix regularization can, in principle, be ``generally
covariant'' if the sum over all background configurations is taken into
account.  A permutation of D-instantons is a discrete analogue of
coordinate reparametrization.

Matrix models pack too much degrees of freedom into a few matrices. As
is often the case, this obscures the meaning of large $N$
limit. Furthermore, in a theory with $T$-duality it is difficult to make
distinction between IR and UV limits. The limits explicitly depend on
the effective dynamics we are talking about. At any rate, it is obvious
in Matrix theory that universal behavior is expected only in large $N$
limit.

We initiated a preliminary study of what class of large $N$ limit is
possible in order to reproduce a QFT. Key idea is to classify the degree
of divergence in terms of $N$, the only source of divergence in Matrix
theory.  It is now possible to interpret renormalization group \`{a} la
Br\'{e}zin and Zinn-Justin \cite{BZ,Hikami,Higuchi} in terms of the
usual renormalization of Yang-Mills theory. We hope to report on this
elsewhere.

One might well be puzzled by the interpretation of the Matrix dynamics
as $d$-dimensional Yang-Mills theory. If the background D-instantons are
distributed really uniformly, the rank of the gauge group would be just
a matter of choice because it depends how we cut the D-instanton gas
into pieces.  Let $\rho$ be the D-instanton density.  Suppose we decide
to call D-instantons inside a $d$-dimensional hypercube of size $b$ as a
cluster.  Then we have
\begin{equation*}
 \begin{array}{lcl}
  n = \rho b^{d},  & \quad &(\mbox{rank of the gauge group}) \\
  \rho' = \rho/n = b^{-d},  && (\mbox{cluster density}) \\
  g_{d}^2 = (2\pi )^{d}\rho' g^{2}=(2\pi )^{d}g^{2}b^{-d}.
   && (\mbox{Yang-Mills coupling}) 
 \end{array}
\end{equation*}
Note that $b$ dependence cancels in the 't Hooft coupling $\lambda
\equiv g_{d}^{2}n = (2\pi )^{d}g^{2}\rho$.  Therefore, matrix
perturbation theory suggests that \emph{any universal property of $U(n)$
gauge theory with adjoint matters should depend, not separately on
$g_{d}$ or $n$, but on 't Hooft coupling $\lambda \equiv g_{d}^{2} n$,
at least for sufficiently small $\lambda$}.

At present, we do not know whether this is generally true or not, but
the following evidence should be taken seriously.  Consider a
renormalization group beta function for $d=4$ Yang-Mills coupling,
\begin{equation*}
 \beta(g_{4}) = - \beta_{0} g_{4}^{3} - \beta_{1} g_{4}^{5} - \cdots.
\end{equation*}
In a $U(n)$ gauge theory with adjoint matters ($C_{2}(G)=n$), the
coefficients are given by $\beta_{0}=c_{0}n,\;\beta_{1}=c_{1}n^{2},\;
\ldots$ with $c_{0},c_{1},\ldots$ depending only on the matter contents.
Thus we have
\begin{equation*}
 \beta(g_{4})= - c_{0} \, n \, g_{4}^{3}- c_{1} \, n^2 g_{4}^{5} -
  \cdots.
\end{equation*}
This can be rewritten as
\begin{equation*}
 \beta(\lambda)= - 2(c_{0} \lambda^{3} + c_{1} \lambda^{5} + \cdots).
\end{equation*}
The right-hand side is a function of $\lambda$ only, in accordance with
our expectation.

\vskip 8mm

This work is supported in part by the Grant-in-Aid for Scientific
Research on Priority Area 707 ``Supersymmetry and Unified Theory of
Elementary Particles'', Japan Ministry of Education.

\end{document}